%% file: nash_babar.tex
\documentclass{ws-p10x7}
\input babarsym
\input symbolspsikst
\input DDsymbols

\input Definitions

\def\Dstarp   {\ensuremath{D^{*+}}\xspace}
\def\Dstarm   {\ensuremath{D^{*-}}\xspace}

\def\DorDstarm {\ensuremath{D^{(*)-}}\xspace}
\def\DorDstarzb  {\ensuremath{\Dbar^{(*)0}}\xspace}
\def\bkg    {\ensuremath {\B \to \Kstar \gamma}}
\newcommand{\rhoppim}{\mbox{$\Bz \to \rho^{\mp} \pi^{\pm}$}}
\newcommand{\rhozpiz}{\mbox{$\Bz \to \rho^0 \pi^0$}}

\newcommand{\bpjpsikp}{\ensuremath{B^+ \to \jpsi K^+}}
\newcommand{\bppsitwoskp}{\ensuremath{B^+ \to \psitwos K^+}}
\newcommand{\bpchiconekp}{\ensuremath{B^+ \to \chic{1} K^+}}
\newcommand{\bzjpsikstarz}{\ensuremath{B^0\to \jpsi K^{*0}}}
\newcommand{\bpjpsikstarp}{\ensuremath{B^+ \to \jpsi K^{*+}}}

\newcommand{\bzjpsikz}{\ensuremath{B^0 \to \jpsi K^{0}}}
\newcommand{\bzjpsipiz}{\ensuremath{B^0 \to \jpsi\piz}}
\newcommand{\bzpsitwoskz}{\ensuremath{B^0 \to \psitwos K^{0}}}

\newcommand{\bzchiconekz}{\ensuremath{B^0 \to \chic{1} K^{0}}}
\newcommand{\bzchiconekstarz}{\ensuremath{B^0 \to \chic{1}K^{*0}}}

\newcommand{\btodsdsk}{\ensuremath{B^+ \to D^{*-} D^{*+} K^+}}
\def\bpsikpm{\ensuremath{\Bpm \to \jpsi \Kpm}}
\def\bpsipipm{\ensuremath{\Bpm \to \jpsi \pipm}}

\def\BRpsipipm  {\ensuremath{\BR(\bpsipipm)}}
\def\BRpsikpm  {\ensuremath{\BR(\bpsikpm)}}
\providecommand{\bfemsix}{${\cal B}(\times10^{-6}$)}

\begin{document}

\title{\babar\ B Decay Results}

\author{Jordan Nash}

\address{Blackett Laboratory,\\
Imperial College,\\
South Kensington,\\
LONDON SW7 2BW, United Kingdom\\
E-mail: j.nash@ic.ac.uk\\
on behalf of the \babar\ collaboration}

\twocolumn[\maketitle\abstract{Data from the first run of the
\babar\ detector at the PEP II accelerator are presented.  Measurements
of many rare B decay modes are now possible using the large data sets
currently being collected by \babar.  An overview of analysis techniques
and results on data collected in 2000 are described. }]

\section{Introduction}
The \babar\ detector\cite{bbrnim} began collecting data just over two years ago with
the heart of our physics program centered around the search for CP
violation in the B meson system.  The excellent performance of the
PEP II accelerator has allowed us to establish the existence of CP
violation in B decays as Jonathan Dorfan has shown today\cite{dorfan,sin2bprl}, 
and also seen by our colleagues at KEK\cite{olsen}.
In searching for CP violation, it is necessary to perform a series
of measurements of rare B decay modes which establish the ability
of the detector to accurately determine the parameters of CP violation.
These measurements also provide an opportunity to look for rare
decays which could give a hint of physics beyond the Standard Model.  In
addition, many of the modes which are now rare modes will provide independent
measurements of CP violating effects during the high luminosity era of
the B factories.

This talk will provide a taste of the physics program which is just
now starting at \babar.  The results presented here are based on
the first Run of \babar\ which lasted until December 2000 and
collected $20.7/fb$ of data on the peak of the \FourS\ resonance
(approximately 22.7 million $B\overline B$ pairs),
and $2.6/fb$ data off-peak.  Most results are preliminary.

\section{B Decays with Charmonium}
The majority of the modes which went into the measurement of \stwob\ 
include charmonium in the final state.  An important first step
was to improve measurements of B mesons into final states including
charmonium, as well as establishing measurements of the B into
previously unseen modes.  
\subsection{Inclusive Cross Sections}\label{subsec:ccincl}
\begin{figure}
\resizebox{0.45\textwidth}{!}{%
\includegraphics{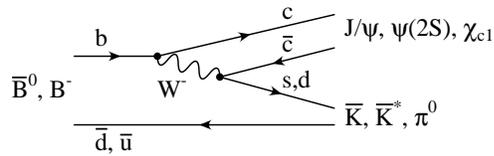}
}
\caption{Feynman diagram for B meson decays to charmonium.}
\label{fig:psifey}
\end{figure}
\begin{figure}
\resizebox{0.45\textwidth}{!}{%
\includegraphics{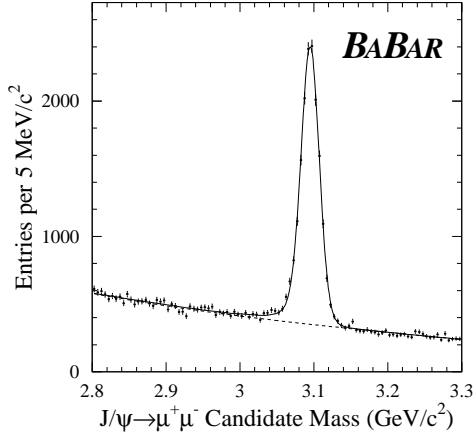}
}
\caption{Inclusive $J/\Psi$ decays into muons.}
\label{fig:psimumu}
\end{figure}

Charmonium is produced in B decays primarily through the
internal spectator diagram shown in figure~\ref{fig:psifey}.
The $J/\psi$ is detected in the leptonic decay channels to
electrons and muons.  An example of the inclusive measurement
of charmonium production is shown in figure~\ref{fig:psimumu}.
The inclusive branching fractions for B decays to 
final states including charmonium are determined to be 
${\cal B}(B\rightarrow J/\Psi X)=(1.044\pm0.013\pm0.028)\times 10^{-2}$,
${\cal B}(B\rightarrow \psitwos X)=(0.275\pm0.020\pm0.029)\times 10^{-2}$,
${\cal B}(B\rightarrow \chic{1} X)=(0.378\pm0.034\pm0.026)\times 10^{-2}$.
In addition, we have also observed the production of charmonium
in the continuum at a rate
$\sigma_{e^+e^-\rightarrow J/\Psi X}=2.52\pm0.21\pm0.21\,pb$\cite{psicont}
which is the first such measurement.

\subsection{Exclusive Cross Sections}\label{subsec:ccexcl}
In order to determine the exclusive cross sections for B decays
we use two main kinematic variables which take advantage of the
fact that the B mesons are produced nearly at rest in the 
\FourS\ rest frame, and that the beam energy is well determined.

The energy-substituted mass
$\mes \equiv \sqrt{{E_{Beam}^{*2}} - {p_B^*}^2}$ uses the beam
energy and the reconstructed B momentum to form an effective mass
for the B candidate.  The mass resolution for $\mes$ is approximately
$2-3\mev$ and the particle mass hypothesis is not needed.  B events
should peak at the B mass in this variable.  
$\DeltaE \equiv E_{B}^* - E_{Beam}^* $ is an orthogonal variable
which takes into account the particle mass hypothesis.  It has
a resolution of approximately $10-20\mev$ depending on the decay
mode, and should be centered at 0 if the particle hypotheses are
correct.  These two kinematic variables are used in nearly all
of our exclusive B decay analyses.

\begin{figure}
\resizebox{0.45\textwidth}{!}{%
      \includegraphics{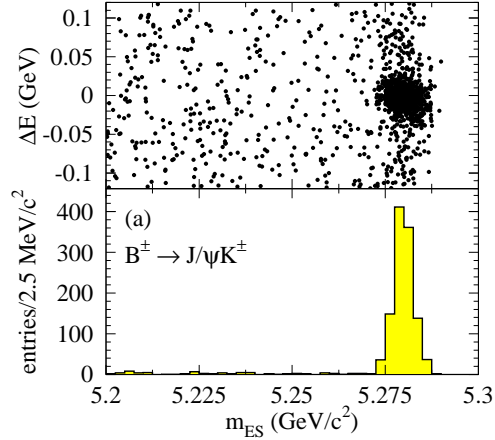}}
\caption{$\DeltaE$ vs $\mes$ for the decay $\bpjpsikp$ the bottom
figure shows the projection of $\mes$ where a cut has been made
on $\DeltaE$.}
\label{fig:psi2d}
\end{figure}

\begin{figure}
\resizebox{0.45\textwidth}{!}{%
      \includegraphics{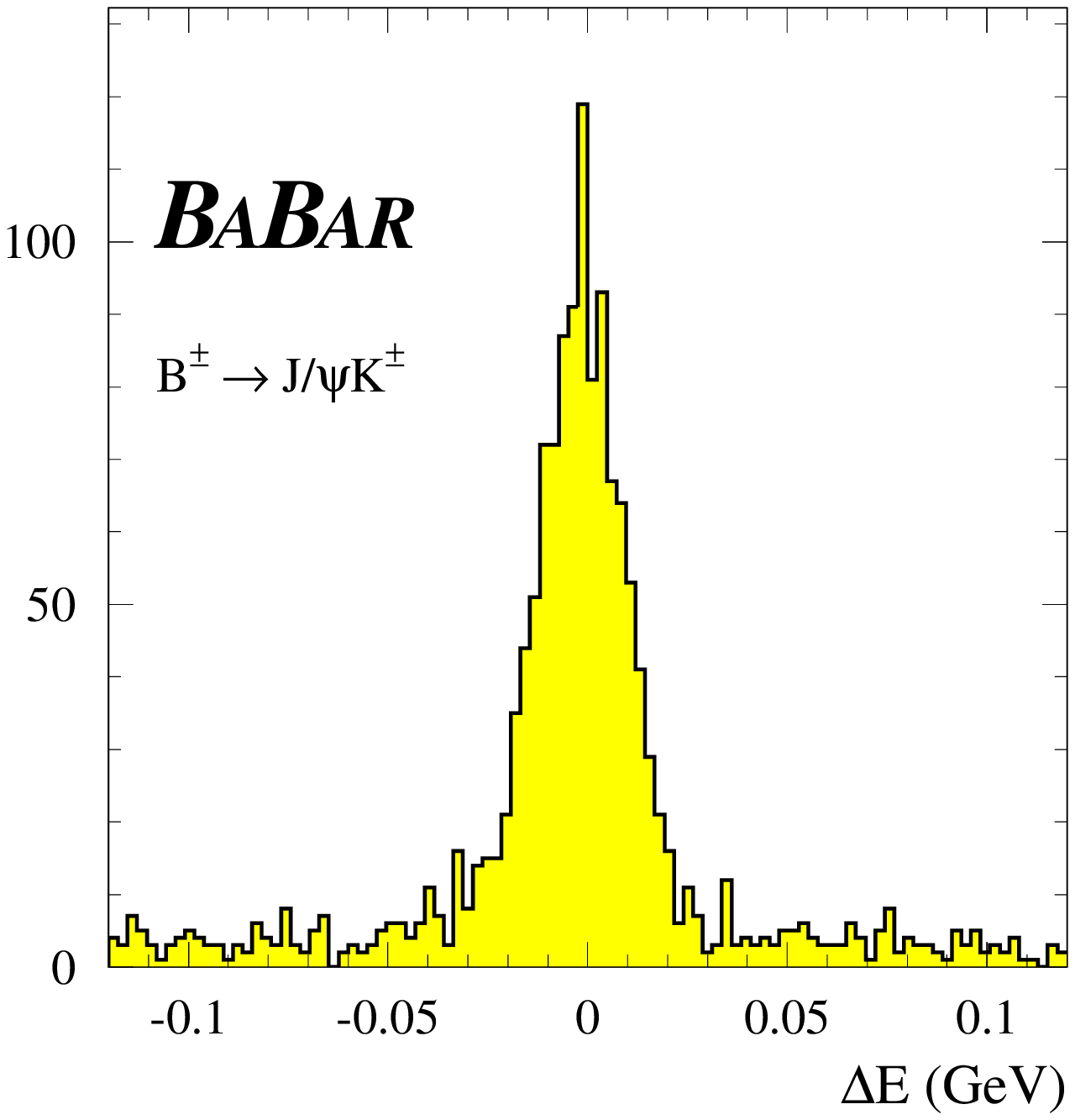}}
\caption{$\DeltaE$ projection for the decay $\bpjpsikp$.}
\label{fig:pside}
\end{figure}

Figure~\ref{fig:psi2d} shows $\DeltaE$ plotted against
$\mes$ for the decay mode $\bpjpsikp$  A selection is made for
$\DeltaE$ centered around 0, and the projection is shown in the
bottom of the figure for the variable $\mes$.  There is a
clear signal with very low background.  The sidebands in $\mes$
are used to estimate the backgrounds under the B peak which are
small in most of the decay modes.  Figure~\ref{fig:pside} shows
an example of the projection of $\DeltaE$ for one mode.

Exclusive branching fractions for many B decay modes with charmonium
in the final state
have been measured \cite{exclcharm}
and are summarized in table~\ref{tab:charmbf}.
In most of these modes, the backgrounds are small and are estimated
using the sidebands.  Backgrounds are primarily from other B decays
which contain charmonium, and these cross-feeds are estimated using
the simulation.

In addition to these branching fraction measurements, we also present
a measurement of the ratio
$\BRpsipipm/\BRpsikpm =[3.91\pm 0.78\pm 0.19]\%$\cite{psikpsipi}, which
significantly improves upon previous measurements, and 
is in good agreement with theoretical predictions.
\begin{table*}
\caption{Branching fraction results for B decays to final states containing charmonium.}
\label{tab:charmbf}
\centerline{
\begin{tabular}{llccccl}
\hline \hline
Channel &   & \multicolumn{5}{c}{Branching fraction/$10^{-4}$}\\
\hline
\bzjpsikz         &  \KS \to \pipi    & 8.5&$\pm$&0.5&$\pm$&0.6 \ \\
                  &  \KS \to \piz\piz & 9.6&$\pm$&1.5&$\pm$&0.7  \ \\ 
                  &  \KL              & 6.8&$\pm$&0.8&$\pm$&0.8   \ \\ 
                  &  All              & 8.3&$\pm$&0.4&$\pm$&0.5  \ \\ 
\bpjpsikp         &                   & 10.1&$\pm$&0.3&$\pm$&0.5 \ \\ 
\bzjpsipiz        &                   & 0.20&$\pm$&0.06&$\pm$&0.02 \ \\ 
\bzjpsikstarz     &                   & 12.4&$\pm$&0.5&$\pm$&0.9   \ \\  
\bpjpsikstarp     &                   & 13.7&$\pm$&0.9&$\pm$&1.1   \ \\  
\bzpsitwoskz      &                   & 6.9&$\pm$&1.1&$\pm$&1.1  \ \\  
\bppsitwoskp      &                   & 6.4&$\pm$&0.5&$\pm$&0.8  \ \\  
\bzchiconekz      &                   & 5.4&$\pm$&1.4&$\pm$&1.1  \ \\  
\bpchiconekp      &                   & 7.5&$\pm$&0.8&$\pm$&0.8  \ \\  
\bzchiconekstarz  &                   & 4.8&$\pm$&1.4&$\pm$&0.9   \ \\  
\hline \hline
\end{tabular}
}
\end{table*}

\subsection{Angular Analysis of \bpsikst}\label{subsec:angular}
One exclusive mode deserves a little more discussion.
The decay mode \bpsikst\ is used to measure \stwob; however
as this mode contains a mixture of odd and even CP amplitudes,
it is necessary to determine the relative contribution of odd and
even CP final states.  This can be accomplished by the use of
an angular analysis of the decay.  

\begin{figure}
\resizebox{0.45\textwidth}{!}{%
\includegraphics{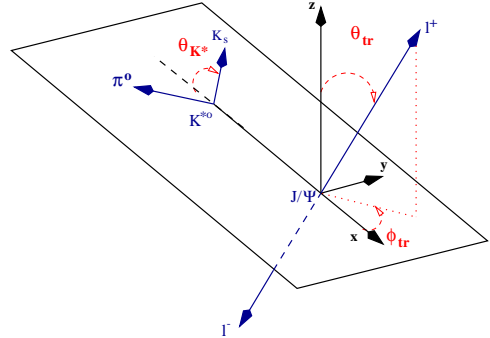}}
\caption{Definition of the angles used to extract the relative
amplitudes in the decay \bpsikst.}
\label{fig:transang}
\end{figure}

\begin{figure}
\epsfxsize120pt
\resizebox{0.45\textwidth}{!}{%
\includegraphics{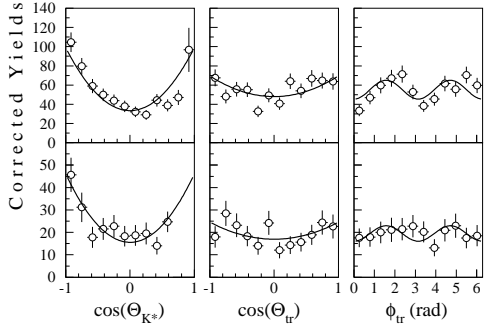}}
\caption{Fit to angular distributions in the decay \bpsikst\ which allow
the extraction of the odd and even CP amplitudes in the decay.}
\label{fig:trans}
\end{figure}

\babar\ has used a transversity
analysis, looking at the decay angles of the \Kstar and the \jpsi\cite{psikstar}.
Figure~\ref{fig:transang} shows the angles used to unfold the relative
amplitudes.
The angular distributions are shown in figure~\ref{fig:trans}, and the results for
${\rm CP}=+1$ states ($\azd$ and $\apd$) and ${\rm CP}=-1$ state ($\atd$) are
tabulated in table~\ref{tab:trans}.

The presence of a significant amplitude of the ${\rm CP}=-1$ state dilutes the
measured CP in this state by an amount $D=1-2|\at|^2$.  The value
measured for $\atd$ implies the dilution factor $D=0.68\pm0.10$.  This
factor is used when the \bpsikst\ are included in the \stwob\ analysis.

\begin{table}
\caption{Results for the odd and even CP amplitudes in the decay \bpsikst.}
\label{tab:trans}
\centerline{\resizebox{0.45\textwidth}{!}{
\begin{tabular}{cllll}  \hline\hline
        Quantity &      \hspace{2.0cm}          &               &       Value                   \\ \hline
        \azd     &                              &       0.597   & $\pm$ 0.028 & $\pm$ 0.024     \\      
        \atd     &                              &       0.160   & $\pm$ 0.032 & $\pm$ 0.014     \\
        \apd     &                              &       0.243   & $\pm$ 0.034 & $\pm$ 0.017     \\ \hline
        	\phit (rad)    	&		&       $-0.17$ & $\pm$ 0.16  & $\pm$ 0.07      \\
        	\phip (rad)   	&              	&       2.50    & $\pm$ 0.20  & $\pm$ 0.08      \\ \hline
\end{tabular}
}}
\end{table}

\section{B Decays with Open Charm}
Decays involving open charm ($D^{(*)}$) allow measurement of either the
Cabibbo-allowed $b\ra ccs$ decay or the Cabibbo-suppressed
$b\ra ccd$ decay.   The latter provides another opportunity to measure
\stwob which is complementary to the measurement made with the charmonium
modes.
\subsection{$B \rightarrow D^{(*)}\bar D^{(*)}K$ Decays}\label{subsec:bddk}

\begin{figure}
\parbox{0.5\textwidth}{
\resizebox{0.24\textwidth}{!}{%
\includegraphics{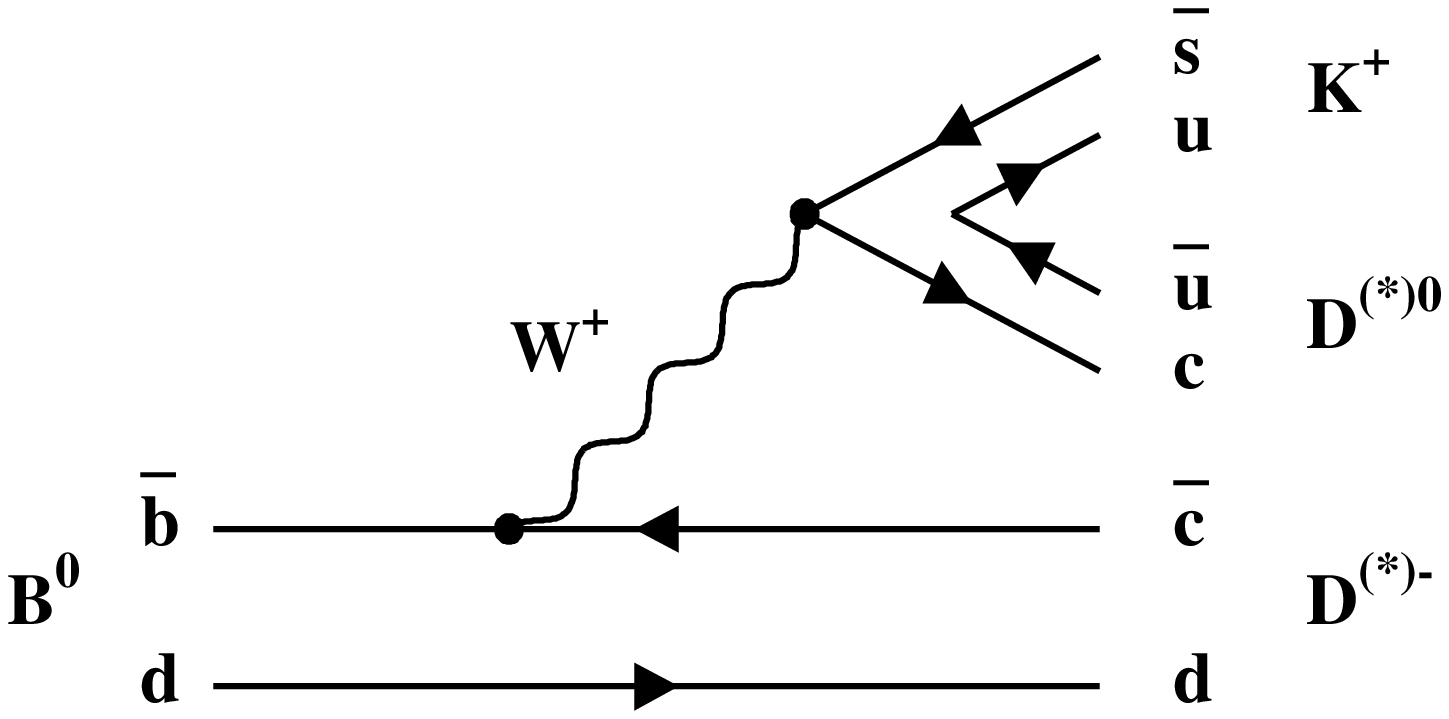}}
\resizebox{0.24\textwidth}{!}{%
\includegraphics{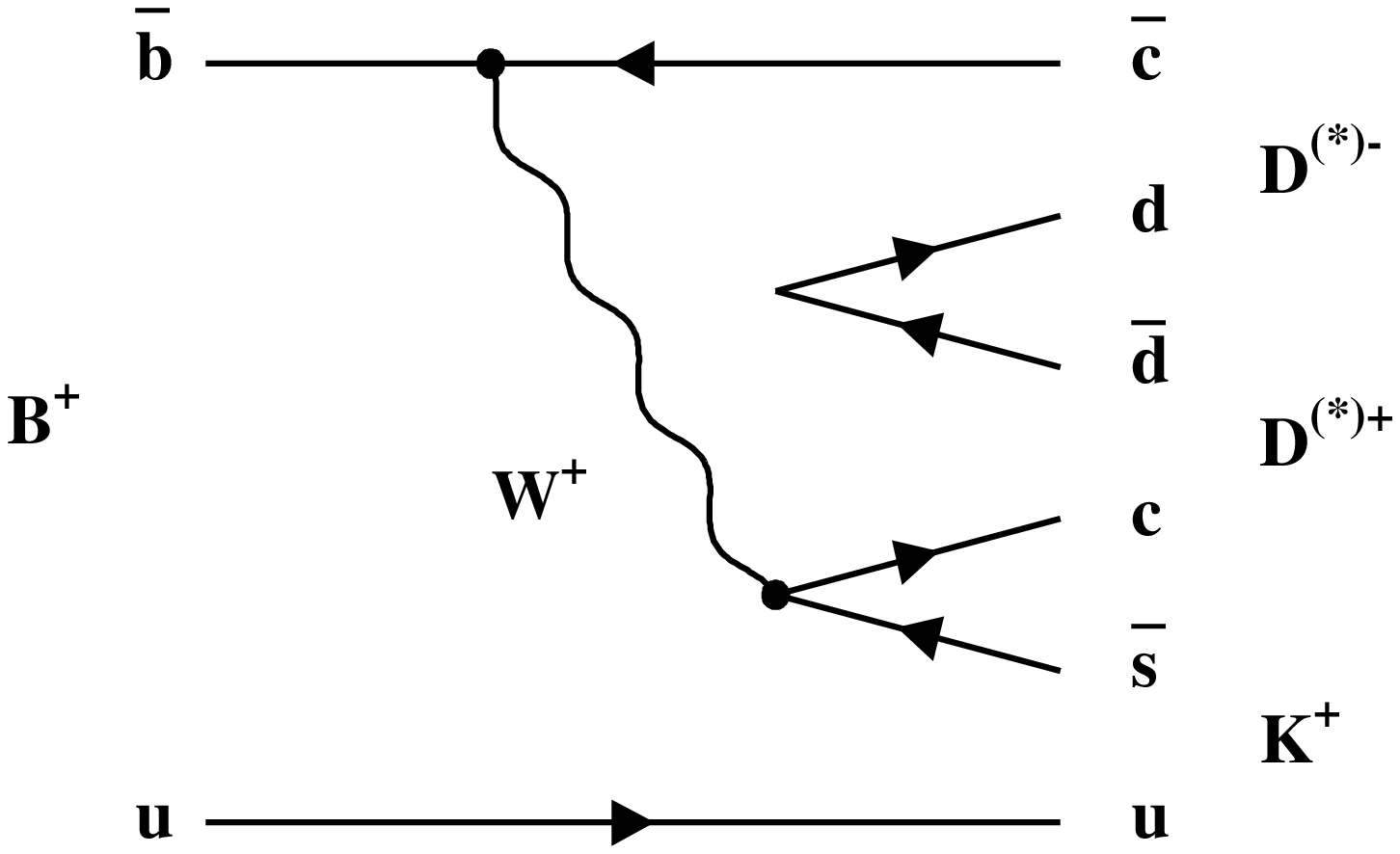}}}
\caption{External (left) and internal (right) spectator diagrams for the decays 
$B \rightarrow D^{(*)}\bar D^{(*)}K$.}
\label{fig:ddkfey}
\end{figure}

In $b\ra ccs$ decays, one expects the $D_s^{(*)+}$ to be the dominant decay
mode.  Previous measurements have however indicated that the 
$B \rightarrow D^{(*)}\bar D^{(*)}K$ decays may have a larger than predicted
contribution to the $b\ra ccs$ rate.   We have looked for the decays
$B \rightarrow D^{(*)}\bar D^{(*)}K$ in both inclusive and exclusive modes\cite{bddk}.  
The decays can occur via both the external diagram shown on
the left in figure~\ref{fig:ddkfey} and the colour-suppressed internal diagram
diagram shown on the right.

\begin{figure}
\resizebox{0.45\textwidth}{!}{%
\includegraphics{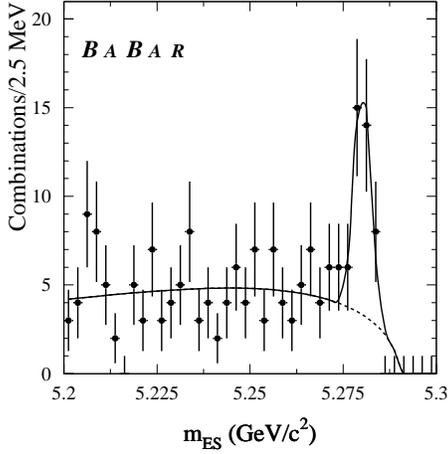}}
\caption{$\mes$ distribution for the decays $\Bz \rightarrow D^{(*)}\bar D^{(*)}K$.}
\label{fig:ddkmes}
\end{figure}

Figure~\ref{fig:ddkmes} shows $\mes$ for all $B^0$ modes.  The 
different decay modes are summed
as there can be several candidates in each event due to the large
number of decay products.  We find for the inclusive branching
fractions
$ {\cal B}(B^0 \ra D^{*-}D^{0}K^+) = (2.8 \pm 0.7 \pm 0.5) \ 10^{-3} $,
$ {\cal B}(B^0 \ra D^{*-}D^{*0}K^+) = (6.8 \pm 1.7 \pm 1.7) \ 10^{-3} $.

\begin{figure}
\resizebox{0.45\textwidth}{!}{%
\includegraphics{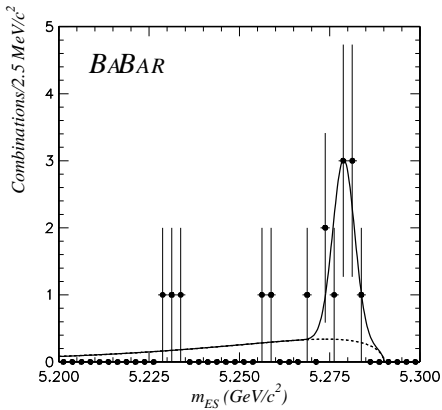}}
\caption{$\mes$ distribution for the exclusive colour-suppressed mode $\Bz \ra D^{*-}D^0K^+$.}
\label{fig:ddkexmes}
\end{figure}

In addition, we have also measured the exclusive mode $\btodsdsk$ for which
the $\mes$ plot is shown in figure~\ref{fig:ddkexmes}.  The branching fraction
is measured to be 
$ {\cal B}(\btodsdsk)= (3.4\pm 1.6\pm 0.9) \ 10^{-3} $.
This is the 
first observation of a colour-suppressed mode not involving charmonium.
\subsection{ \Bztodstdst }\label{subsec:bdsds}

\begin{figure}
\parbox{0.5\textwidth}{
\resizebox{0.24\textwidth}{!}{%
\includegraphics{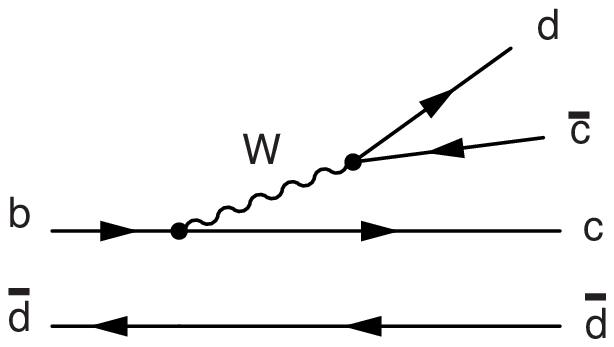}}
\resizebox{0.24\textwidth}{!}{%
\includegraphics{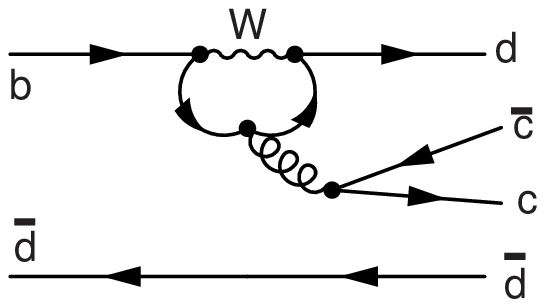}}}
\caption{Tree-level diagram (left) and penguin (right) diagrams contributing to \Bztodstdst decays.}
\label{fig:ddtree}
\end{figure}

Decays to two $D$ mesons proceed via Cabibbo-suppressed diagrams as shown in 
figure~\ref{fig:ddtree}.  
This decay mode is sensitive to \stwob.  However, as in the decay \bpsikst,
this is a vector-vector decay mode, and there are both CP-odd and CP-even
components.  These will need to be measured before \stwob\ can be extracted
from the data.  Again these modes are difficult to reconstruct due to the
large number of particles in the final states.  The $D^*$ modes offer the 
most constraints from the mass difference with the soft pion, and so are
reconstructed with a reduced background.  We have measured the branching
fraction for \Bztodstdst\ \cite{bdstdst} with the next step being the
measurement of the amplitudes of the CP components.

The branching fraction for this mode is Cabibbo-suppressed and should
be given approximately by
$$\approx
\left(\frac{f_{D^{(*)}}}{f_{D_S^{(*)}}}\right) \tan^2\theta_C
{\BR}(\BtoDsDbar)$$
which is of order 0.1\%.
We reconstruct both \Dstptopip,  \Dstptopiz, with the 
$\Dz$ in the decay modes 
$\Km \pip$, $\Km \pip \piz$, $\Km \pip \pip \pim$, $\KS \pip \pim$ 
and the $\Dp$ in the modes
$\Km \pip \pip$, $\KS \pip$, $\Km \Kp \pip$.

\begin{figure}
\resizebox{0.45\textwidth}{!}{%
\includegraphics{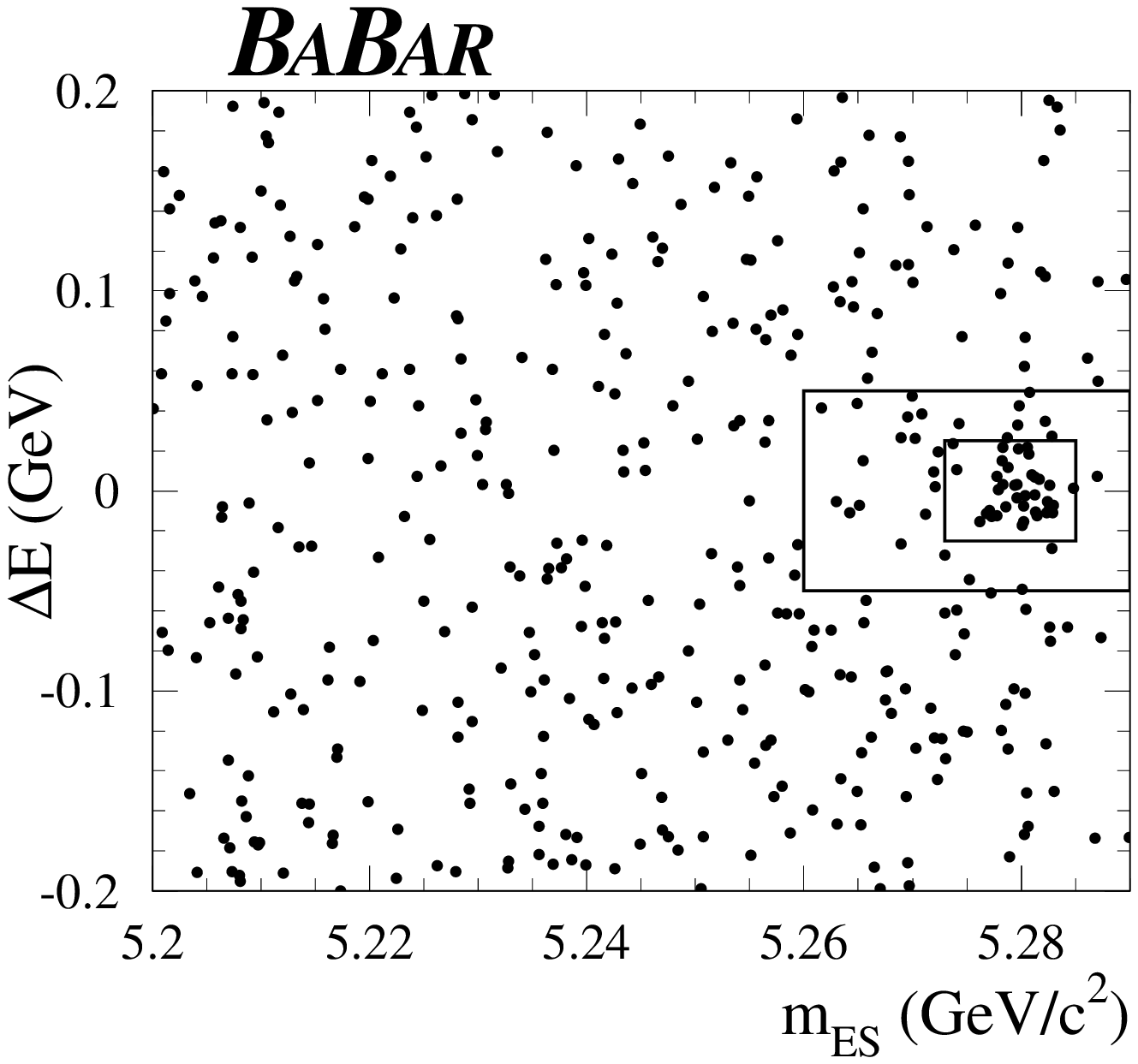}}
\caption{$\mes$ versus $\DeltaE$ for the decay \Bztodstdst.}
\label{fig:ddmesde}
\end{figure}

\begin{figure}
\resizebox{0.45\textwidth}{!}{%
\includegraphics{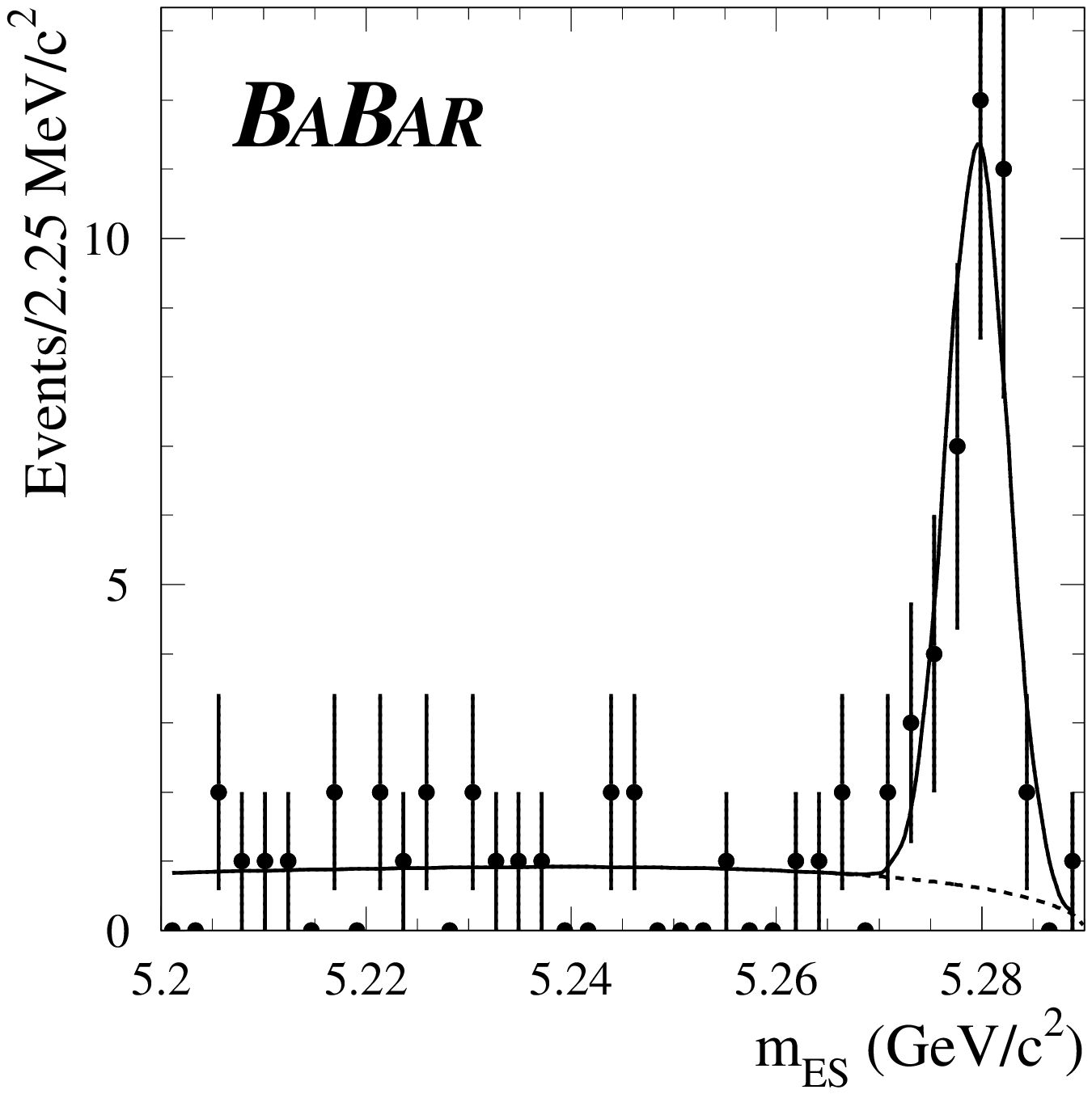}}
\caption{$\DeltaE$ projection for the decay \Bztodstdst.}
\label{fig:ddde}
\end{figure}

Background is determined by looking in the sidebands in 
the $\mes$ vs $\DeltaE$ plot (figure~\ref{fig:ddmesde})
in the regions outside the signal box, and scaling this
by the relative areas of the signal and background regions
to estimate the amount of background in the signal box.
A projection of $\mes$ in the signal box is shown in
figure~\ref{fig:ddde}.  We find a background subtracted
signal of   $31.8 \pm 6.2 \pm 0.4$ events which gives
a branching fraction
$\BRbztodstdst$.  

The main systematic uncertainties in this analysis come
from tracking efficiency (9.4\%), due to the large number
of tracks in the final state, and the fact that the
polarization of the final state is unmeasured (6.6\%).

\section{Charmless B Decays}
One of the most exciting prospects for the coming run will
be the start of the process of measuring $\stwoa$.  Charmless
B decays are where the B factories will perhaps have their
greatest unique advantage in measurements of CKM parameters.
\babar\ have presented the first attempts at these measurements
at this conference\cite{dorfan}, which illustrate the prospects
and the difficulties in carrying out these measurements.

\begin{figure}
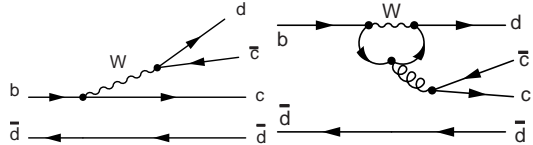

\parbox{0.5\textwidth}{
\resizebox{0.24\textwidth}{!}{%
\includegraphics{plots/DstarFeyn.eps}}
\resizebox{0.24\textwidth}{!}{%
\includegraphics{plots/Penguindiag.eps}}}
\caption{Tree-level diagram (left) and penguin (right) diagrams contributing to charmless B decays.}
\label{fig:clfey}
\end{figure}

One of the main difficulties in these measurements is the fact
that the penguin diagram (figure~\ref{fig:clfey}) contribution
is most likely of comparable magnitude to the tree diagram.
The implication of this is that one must determine the relative
magnitudes of the two contributions in order to accurately determine
the angle $\stwoa$.  One measures an asymmetry which is only an
effective $\stwoa$ and then needs to correct this based on the 
penguin/tree magnitudes.  

There are several strategies for extracting these magnitudes.  
In the two-body modes, it is possible to perform an isospin 
analysis which allows one to determine the relative contribution
of the penguin and tree diagrams, but it requires the difficult
measurements of both $\Bz \to \piz \piz$ and $\Bzb \to \piz \piz$. 
Another strategy involves a full Dalitz plot analysis of the 
three-body decay modes.

Other difficulties in measuring these modes are the small value of
$V_{ub}$ which implies small branching fractions as well
as the fact that these modes suffer from severe combinatoric
backgrounds.  

In addition to measurement of $\stwoa$, charmless modes also
have potential for measurements of direct CP violation as will
be discussed in section~\ref{subsec:directCP}.

\subsection{Two-body Charmless Decays}\label{subsec:cl2body}
\begin{figure}
\resizebox{0.45\textwidth}{!}{%
\includegraphics{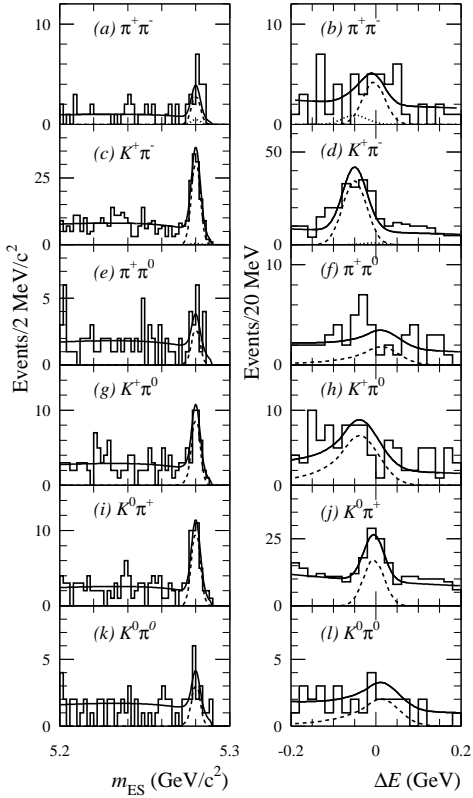}}
\caption{$\mes$ and $\DeltaE$ projections for charmless two-body
B decay modes.}
\label{fig:cl2bmesde}
\end{figure}
Our two-body decay analysis\cite{cl2b} makes use of a 
maximum likelihood fit where the input to the fit includes
a Fisher discriminant based on event shape variables, the
$\mes$ and $\DeltaE$ distributions (shown in figure~\ref{fig:cl2bmesde}),
and the Cherenkov angle measured in the \babar\ DIRC\cite{bbrnim}.
The DIRC is effective at separating pions from kaons at the high
momenta seen in two-body decay modes, and this combined with the
measured $\DeltaE$ provides the ability to distinguish between
$\Bz \to \pip \pim$ and $\Bz \to \Kp \pim$.  Results for the
branching fractions measured, and limits on unobserved modes,
are summarized in table~\ref{tab:cl2bbf}.

\begin{figure}
\resizebox{0.45\textwidth}{!}{%
\includegraphics{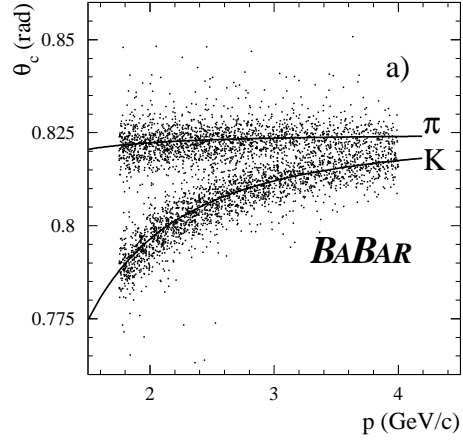}}
\caption{Cherenkov angle measured for pions and kaons as a function of momentum.}
\label{fig:cl2bdirc}
\end{figure}
\begin{table}
\caption{$\mes$ and $\DeltaE$ measurements for two-body
charmless B decays.}
\label{tab:cl2bbf}
\resizebox{0.45\textwidth}{!}{
\begin{tabular}{lcccccc} 
Mode  & $\varepsilon$ (\%) & $N_S$ & $S$ ($\sigma$) & 
\BR($10^{-6}$)\\ 
\hline
$\pip\pim$ &  $45$  & $41\pm 10\pm 7$ & 
$4.7$  & $4.1\pm 1.0\pm 0.7$ \\
$\Kp\pim$ &  $45$ & $169\pm 17\pm 13$ &
$15.8$ & $16.7\pm 1.6\pm 1.3$ \\
$\Kp \Km$  & $43$ & $8.2^{+7.8}_{-6.4}\pm 3.5$  &
$1.3$  & $<2.5$ ($90\%$ C.L.) \\\hline
$\pip\piz$  & $32$ & $37\pm 14\pm 6$ & 
$3.4$  & $<9.6$ ($90\%$ C.L.) \\
$\Kp\piz$    & $31$ & $75\pm 14\pm 7$ & 
$8.0$  & $10.8^{+2.1}_{-1.9}\pm 1.0$ \\
$\Kz\pip$   & $14$ & $59^{+11}_{-10}\pm 6$ &
$9.8$  & 
$18.2^{+3.3}_{-3.0}\pm 2.0$ \\
$\Kzb\Kp$    & $14$ & $-4.1^{+4.5}_{-3.8}\pm 2.3$  &
$-$    & $<2.4$ ($90\%$ C.L.)\\\hline
$\Kz\piz$  & $10$ & $17.9^{+6.8}_{-5.8}\pm 1.9$ &
$4.5$  & $8.2^{+3.1}_{-2.7}\pm 1.2$ \\
$\KzKzb$  & 36  & $3.4^{+3.4}_{-2.4}\pm 3.5$ &
$1.5$  & $<10.6$ ($90\%$ C.L.) \\
\end{tabular}
}
\end{table}
\subsection{Quasi-two-body Charmless Decays}\label{subsec:cl12body}
\begin{figure}
\resizebox{0.45\textwidth}{!}{%
\includegraphics{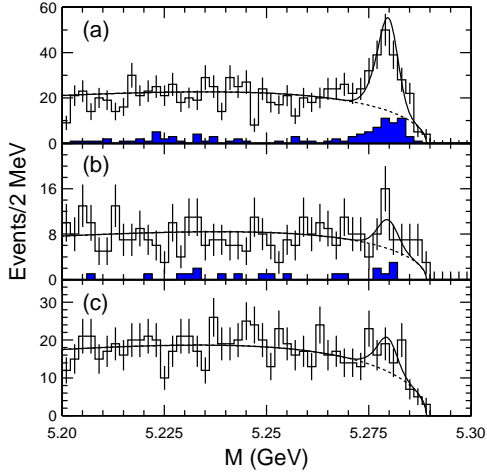}}
\caption{$\mes$ distributions for quasi-two-body decays.  The shaded area is $\eta'\ra\eta\pi\pi$.}
\label{fig:etapk}
\end{figure}

The CLEO result for the decay branching fraction of $\Bp\ra\eta'K^+$\cite{cleoetapk}
was considerably higher than expected from heavy flavour theory\cite{thy}.  We have looked
for this decay mode in the Run~1 data.
The analysis proceeds\cite{etapk} by reconstructing 
$\eta'$ in the modes $\eta'\ra \eta\pi^+\pi^-$ or $\rho^0\gamma$
and $\omega$ in $\omega\ra\pi^+\pi^-\piz$.  The analysis performs
an unbinned maximum likelihood fit on the distributions of
$\Delta E$,$M$,$m_R$, $m_\eta$, and a Fisher discriminant based
on event shape variables.
A significant signal is found in the modes
a) $B^+\ra \eta'K^+$, 
b) $B^0\ra \eta'K^0$, and
c) $B^+\ra \omega\pi^+$ which are shown in figure~\ref{fig:etapk},
and the results are summarized in table~\ref{tab:quasi2b}.
\begin{table}
\caption{Measured branching fractions and upper limits for quasi two-body B decay modes.}
\label{tab:quasi2b}
\begin{tabular}{lcc}
\hline\hline
Mode			& $S$		& \bfemsix\ (90\% CL)\\
\hline
$\eta' K^+$		& 17		& $70\pm 8\pm 5$	\\
$\eta' K^0$		& 5.9		& $42^{+13}_{-11}\pm 4$	\\
$\eta' \pi^+$		& 2.8		& $5.4^{+3.5}_{-2.6}\pm 0.8$\ $(<12)$		\\
\hline                                                            		
$\omega K^+$		& 1.6		& $1.4^{+1.3}_{-1.0}\pm 0.1$ \ $(<4)$	\\
$\omega K^0$		& 3.2		& $6.4^{+3.6}_{-2.8}\pm 0.8 $ \ $(<12)$\\
$\omega\pi^+$		& 5.1		& $6.6^{+2.1}_{-1.8}\pm 0.7 $		\\
$\omega\pi^0$		& $-$		& $-4.6\pm 2.7\pm 1.2 $ \ $(<4)$	\\
\hline\hline
\end{tabular}
\end{table}
\subsection{Branching Fractions \BetaKstz and \BetaKstp }
\begin{figure}
\resizebox{0.45\textwidth}{!}{%
\includegraphics{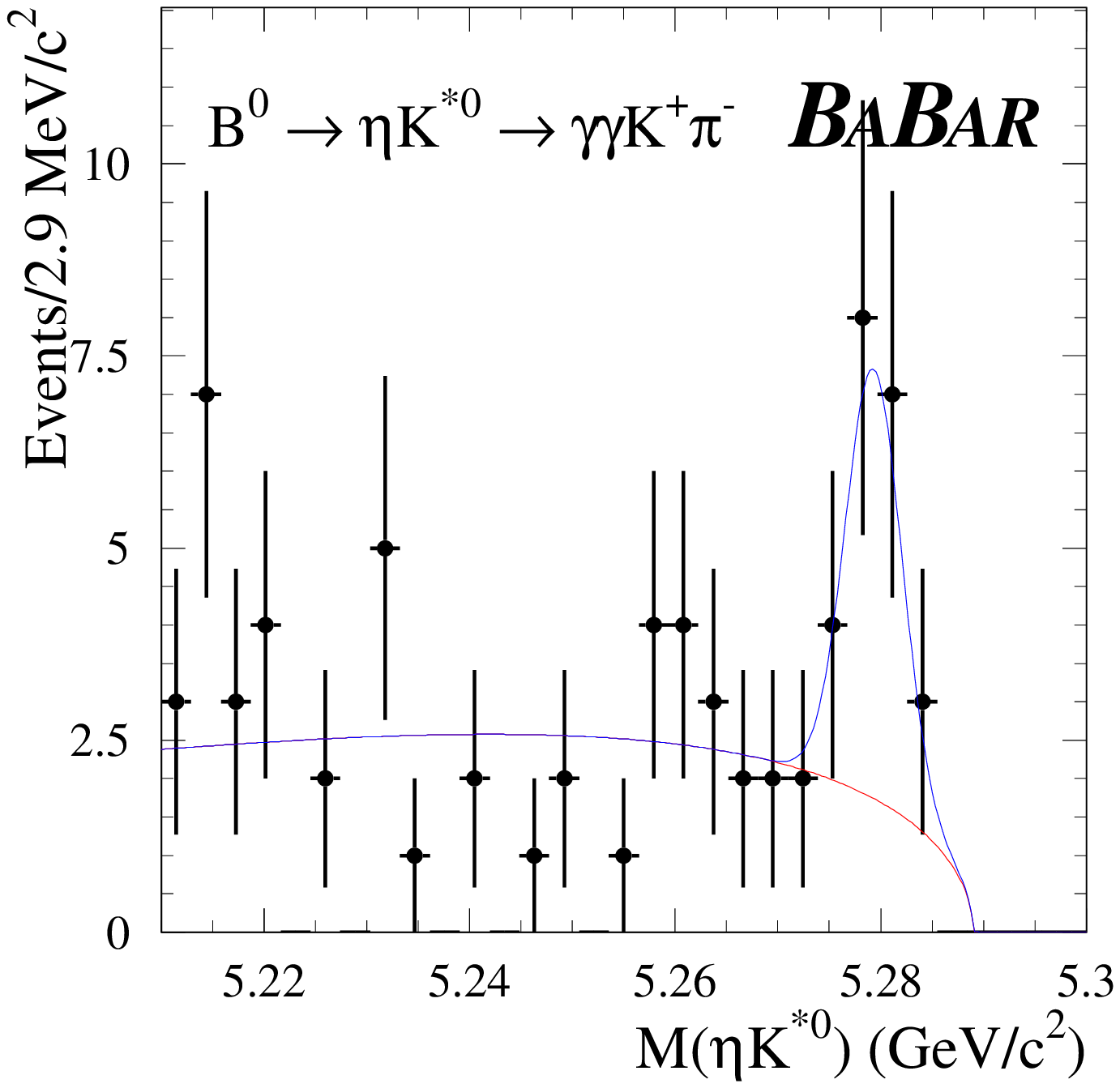}}
\caption{$\mec$ for $B^0\ra\eta K^{*0}$.}
\label{fig:etakst}
\end{figure}

Another set of modes with anomolously high branching fractions\cite{cleoetakst} are
$B^0\ra\eta K^{*0}$ and $B^+\ra\eta K^{*+}$.
These are searched for in the decay modes 
$\eta\ra \gamma\gamma$, $K^{*0}\ra K^+\pi^-$, and $K^{*+}\ra\KS\pi^+$.
The analysis\cite{etakst} uses an unbinned maximum likelihood fit using
$\Delta E$, $M_{\gamma\gamma}$, $M_{K\pi}$, and a Fisher discriminant based
on event shape variables.  The B candidate mass is calculated using a full
kinematic fit for the mass of the B including the beam energy constraint,
$\mec$.   The results for $\mec$ are shown in figure~\ref{fig:etakst}
and the results for the branching fractions are shown in table~\ref{tab:etakst}.
For both this analysis and the analysis in the previous section, we confirm
the higher than expected branching fractions.
\begin{table}
\caption{Branching fractions for $B^0\ra\eta K^{*0}$ and $B^+\ra\eta K^{*+}$.
Also shown are the signal yield and significance.  An upper limit is
also given for $B^+\ra\eta K^{*+}$.}
\label{tab:etakst}
\centerline{
\resizebox{0.45\textwidth}{!}{
\begin{tabular}{lccccc}
\dbline
Mode    & Signal yield      & $S$     & \bfemsix\ & CL 90 \%      \\
\sgline
\quad\fetaKstz   & $21 \pm 6$  &  5.4    & $19.8^{+6.5}_{-5.6} \pm 1.7$ & \\
\quad\fetaKstp    &$14 \pm 7$ & 3.2      & $22.1^{+11.1}_{-9.2} \pm 3.3$ &33.9\\
\dbline
\end{tabular}
}}
\end{table}
\subsection{Evidence for $B^0\to a_0^\pm(980)\pi^\mp$ }
\begin{figure}
\resizebox{0.45\textwidth}{!}{%
\includegraphics{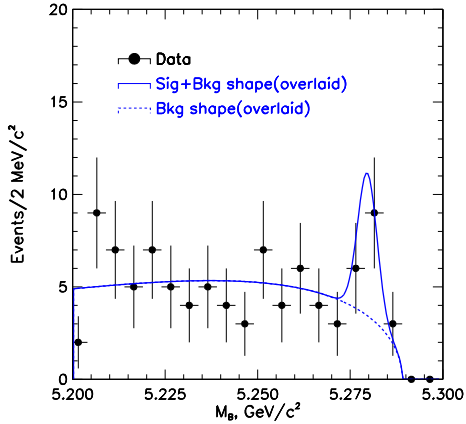}}
\caption{$\mec$ distribution for the decays $B^0\to a_0^\pm(980)\pi^\mp$.}
\label{fig:a0pi}
\end{figure}

The previously unobserved mode $B^0\to a_0^\pm(980)\pi^\mp$ has potential
to provide a measurement of \stwoa\ in a quasi-two-body analysis.  We
search for the decay $B^0\to a_0^\pm(980)\pi^\mp$ 
where $a_0\ra\eta\pi$ and $\eta\ra\gamma\gamma$.  A maximum likelihood
fit is performed to increase the sensitivity to the signal.  The fit
includes $\DE$, $M_\eta$, $\mec$, as well as Fisher and neural network
discriminants based on event shape variables.  The dominant background
in this channel comes from continuum events.   The $\mec$ distribution
for this mode is shown in figure~\ref{fig:a0pi}. The fit gives a
branching fraction measurement with a $3.7\sigma$ significance
with a value\cite{a0pi}
$	{\BR}(B^0\to a_0^\pm 
		(a_0^\pm \rightarrow \eta \pi^\pm) \pi^\mp) 
		\,=\,(6.7_{\,-2.7}^{\,+3.2}\pm1.2)\times 10^{-6}~.
$
This implies an upper limit on the branching fraction
of this mode of ${\BR} < 11.2 \times 10^{-6}$ ( 90\% C.L.).
\subsection{Measurements of $B^0$ Decays to $\pi^+\pi^-\pi^0$}\label{subsec:cl3body}
\begin{figure}
\resizebox{0.45\textwidth}{!}{%
\includegraphics{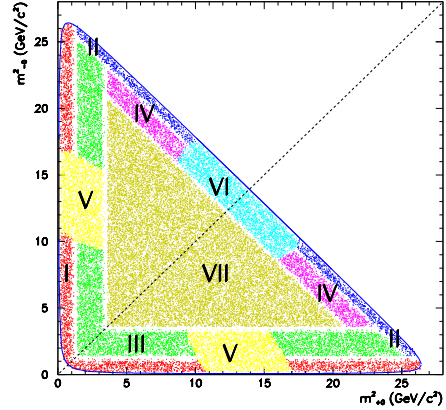}}
\caption{Regions of the three-pion Dalitz plot used for determining branching fractions.}
\label{fig:dalitz}
\end{figure}

The three pion decays of the $\Bz$ offer another method for extracting $\stwoa$ which 
exploits the interference
between the \rhoppim\ modes and the colour-suppressed \rhozpiz.\cite{bbrphys}
These modes suffer from a large combinatoric background coming from continuum events
and small branching fractions.   The current analysis\cite{rhopi} uses a Fisher
discriminant based on 11 shape variables in order to distinguish between signal
and background.   Extraction of $\stwoa$ will require performing an amplitude
analysis in the three-pion Dalitz plot.  This analysis seeks to measure 
contributions to the three-pion branching fractions from different regions
of the Dalitz plot.  The data are sorted into seven samples based on which
area of the Dalitz plot the three-body decay falls into.  Figure~\ref{fig:dalitz}
shows the regions of the Dalitz plot which are used in the analysis.

A significant signal is seen only in the mode $\rhoppim$ (which corresponds
to the regions labelled I in the plot).  Upper limits are set for the branching
fraction in other regions of the Dalitz plot, and are summarized in table~\ref{tab:cl3b}.
\begin{table*}
\caption{Results for the three-pion B decay modes in the regions of the Dalitz plot.  Shown are the number of
signal and background events, the efficiency for the mode, the signifigance of the result, the branching fraction
measurement, and upper limits.  A significant result is seen only in the mode $\Bz\to\rho^+\pi^-$.}
\label{tab:cl3b}
\centerline{
\begin{tabular}{lcccccc}
\hline\hline
\multicolumn{1}{c}{Mode} & Signal  & Bkgd              & $\epsilon$ & Sig.    &${\mathcal B}/10^{-6}$ & ${\mathcal B}/10^{-6}$\\
                         &         & $\qqbar + \bbbar$ &            &$\sigma$ &                       &      90\% C.L. \\\hline
(I)$\Bz\to\rho^+\pi^-$	&  42.8    & 78.2              & 0.13       & 5.0     & $28.9\pm 5.4\pm 4.3$  & \\
(I)$\Bz\to\rho^-\pi^+$	&  46.2    & 71.8              &            &         &                       & \\\hline
(II)$\rhozpiz$		&  6.1     & 20.9              & 0.07       & 1.0     & $3.6\pm 3.5\pm 1.7$    & 10.6  \\
(III)$\rho^\pm(1450)$   &  17.4    & 57.6              & 0.15       & 1.8     & $5.1\pm 2.9\pm 2.2$    & 11.3 \\
(IV)$\rho^0(1450)$	&  -4.7    & 12.7              & 0.09       &         &                       & 2.7 \\
(V) charged Scalar      &  8.6     & 35.4              & 0.15       & 0.4     & $2.5\pm 2.1\pm 0.8$    & 6.1  \\
(VI) $f^0$              &  -0.3    & 6.3               & 0.07       &         &                       & 5.2 \\
(VII)$(NR)$             &  -4.2    & 45.2              & 0.07       &         &                       & 7.3 \\\hline\hline
\end{tabular}
}
\end{table*}
\section{Other Rare B Decays}
Rare B decay modes which proceed primarily through penguin or higher-order 
weak transitions provide an opportunity to search for influences
of non-Standard Model processes.  These can show
up as larger than expected cross sections, or possibly in direct
CP-violating effects.  With the high luminosity of the B factories
it is now possible to begin searching for modes with branching
fractions at the $10^{-7}$ level.
\subsection{$B^0 \rightarrow \gamma \gamma$ }\label{subsec:rarebgg}
\begin{figure}
\parbox{0.45\textwidth}{
\centerline{
\resizebox{0.23\textwidth}{!}{%
\includegraphics{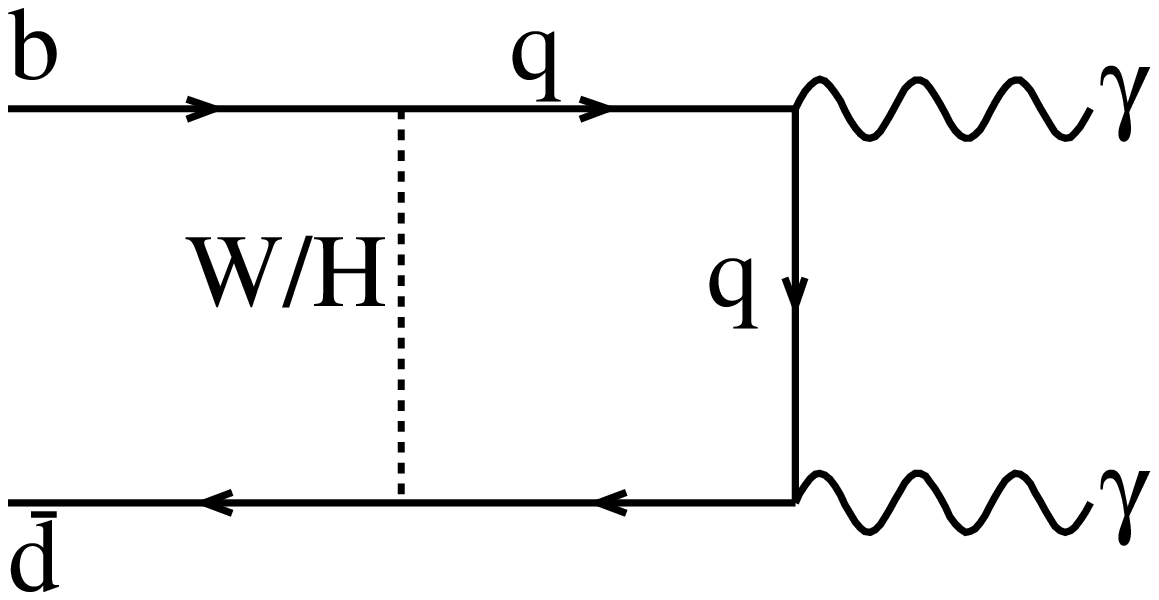}}}
\centerline{
\resizebox{0.23\textwidth}{!}{%
\includegraphics{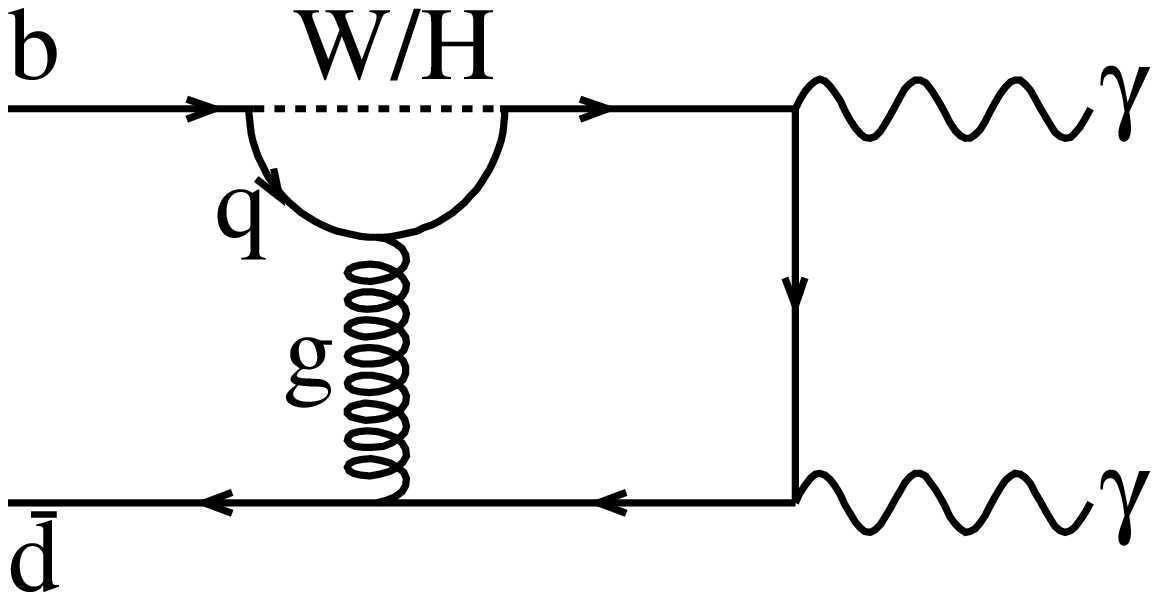}}}}
\caption{SM Diagrams for $B \rightarrow \gamma \gamma$.}
\label{fig:bggfey}
\end{figure}

\begin{figure}
\resizebox{0.45\textwidth}{!}{%
\includegraphics{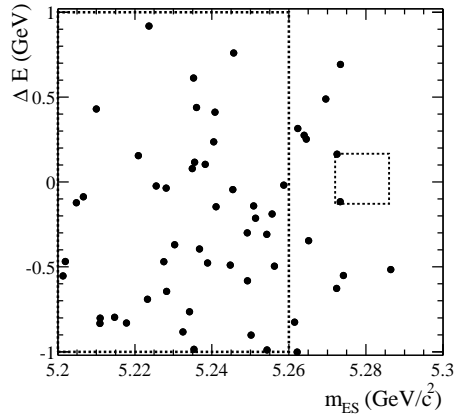}}
\caption{$\mes$ vs $\DeltaE$ for $B \rightarrow \gamma \gamma$.}
\label{fig:bggmesde}
\end{figure}

The decay $B^0 \rightarrow \gamma \gamma$ is expected to occur
in the Standard Model at a branching fraction of approximately
$10^{-8}$, where the contributing diagrams are shown in figure~\ref{fig:bggfey}.
We look for these events\cite{bgamgam}
by searching for two high energy
photons, where at least one photon has $2.3 < E^*_\gamma < 3.0 \gev$.
Photons which can be combined with another photon to create a 
$\piz$ candidate are rejected.  The $\mes$ versus $\DeltaE$ plot
for events which pass all selections is shown in figure~\ref{fig:bggmesde}.
One candidate event lies within the signal box, and we set a branching
fraction upper limit
${\cal B}(B^0 \rightarrow \gamma \gamma ) < 1.7 \times 10^{-6}$, which
is more than an order of magnitude improvement on the previous limits.
\subsection{$B \rightarrow K \ell^+ \ell^-$ and  $ B \rightarrow K^{\ast}(892) \ell^+ \ell^-$}\label{subsec:rarekll}
\begin{figure}
\resizebox{0.45\textwidth}{!}{%
\includegraphics{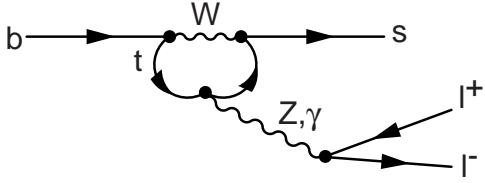}}
\caption{Diagram for $B \rightarrow K \ell^+ \ell^-$.}
\label{fig:kllfey}
\end{figure}

The decay $B \rightarrow K \ell^+ \ell^-$, which in the Standard Model proceeds
via the diagram shown in figure~\ref{fig:kllfey}, is predicted\cite{kllth} to occur at a branching
fraction of order $10^{-7}-10^{-6}$.  These rates are now
becoming accessible at the B factories, and we should expect to determine whether
the Standard Model predictions are valid.  

\begin{figure}
\resizebox{0.45\textwidth}{!}{%
\includegraphics{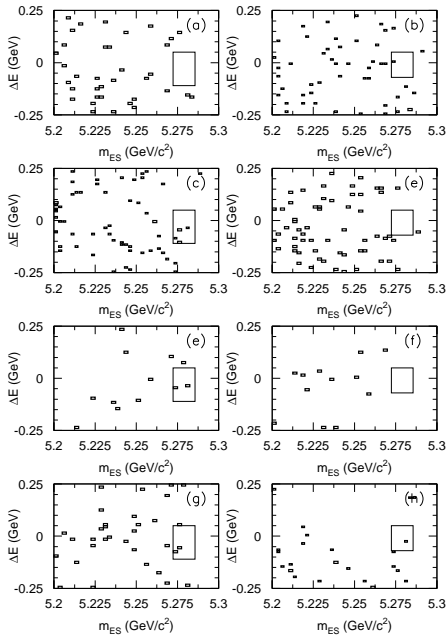}}
\caption{$\Delta E$ vs.~$m_{\rm ES}$ for all modes $B \rightarrow K \ell^+ \ell^-$ and $ B \rightarrow K^{\ast} \ell^+ \ell^-$.  The
modes are labelled in table~\ref{tab:kll}.}
\label{fig:kllmes}
\end{figure}

It is vital to control the backgrounds in this mode as the 
copious production of $B\to J/\psi K^{(*)}$ mimics the signal.  
Events where the invariant mass of the two leptons is consistent
with a $J/\psi$ are vetoed.  Both the signal and sideband regions
for this analysis were kept hidden during the definition of the event
selection procedure to avoid bias.  
The $\mes$ vs $\DeltaE$  plots for the eight modes studied are
shown in figure~\ref{fig:kllmes}.  No evidence for a signal is
seen in any of the modes, and we set upper limits of 
${\mathcal B} (B\to K\ell^+\ell^-)  < 0.6\times 10^{-6}  \  {\rm at}\ 90\%  \ {\rm C.L.} $ and
${\mathcal B} (B\to K^*\ell^+\ell^-)  < 2.5\times 10^{-6} \   {\rm at}\ 90\%  \ {\rm C.L.} $
which are close to the Standard Model predictions.  One could anticipate
seeing a signal in this decay mode with the data from the next run of \babar\
if the Standard Model calculations are correct.  A summary of the data is
shown in table~\ref{tab:kll}.

\begin{table*}
\caption{Branching fraction upper limits for the modes $B \rightarrow K \ell^+ \ell^-$ and $ B \rightarrow K^{\ast} \ell^+ \ell^-$.
Also shown are the fitted signal yields, and expected background in each of the modes and the efficiency.}
\label{tab:kll}
\centerline{
\begin{tabular}{lcccccc}
\hline\hline
\multicolumn{1}{c}{Mode}                & Signal  & 90\% CL &  Equiv. & $\epsilon$  & ${\mathcal B}/10^{-6}$  & ${\mathcal B}/10^{-6}$ \\
                                        &  yield      & yield     &  bkg. &  (\%)    &   &      90\% CL  \\
 \hline 
(a)$K^+e^+e^-$                      &  -0.2  &  3.0      &   0.6  & 17.5 & -0.1       &  0.9 \\    
(b)$K^+\mu^+\mu^-$                  &  -0.2  &  2.8      &   0.4  & 10.5 &-0.1      &  1.3  \\
(c)$K^{*0}e^+e^-$                   &   2.5  &  6.7      &   1.8  & 10.2 &1.6      &  5.0  \\
(d)$K^{*0}\mu^+\mu^-$               &  -0.3  &  3.6      &   1.1  & 8.0  &-0.2      &  3.6 \\
(e)$K^0 e^+e^-$                   &   1.3  &  5.0      &   0.3  & 15.7   &1.1      &  4.7  \\
(f)$K^0 \mu^+\mu^-$               &  0.0 &  2.9      &   0.1  & 9.6     &0.0       &  4.5 \\
(g)$K^{*+}e^+e^-$           &   0.1 &  3.8      &   0.9  & 8.5         &0.1        &  10.0 \\
(h)$K^{*+}\mu^+\mu^-$       &   1.0  &  4.3      &   0.5  & 5.8        &3.3        &  17.5 \\
\hline \hline
\end{tabular}
}
\end{table*}
\subsection{$\B \to \Kstar \gamma$ Branching Fractions }\label{subsec:rareksg}
$\B \to \Kstar \gamma$ proceeds through a penguin diagram
similar to that in figure~\ref{fig:kllfey}.  This mode has the
potential to be sensitive to the presence of SUSY or charged Higgs
contributions.   We have measured the branching fraction into
four exclusive modes.\cite{kstargam}  

\begin{figure}
\resizebox{0.45\textwidth}{!}{%
\includegraphics{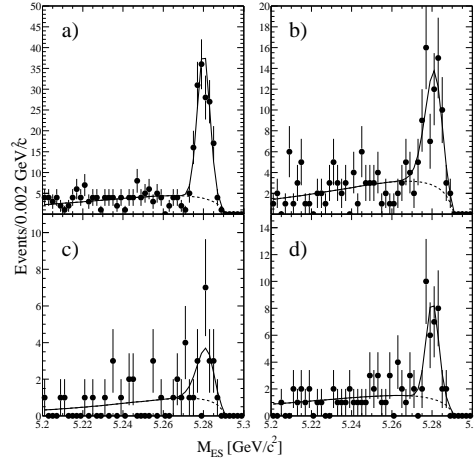}}
\caption{$\mes$ for each of the $\B \to \Kstar \gamma$ decay modes.
The modes are identified in table~\ref{tab:kstgam}.}
\label{fig:kstgam}
\end{figure}
The main background in this analysis comes from 
$e^+e^-\ra q\overline{q}\gamma$ and
$e^+e^-\ra q\overline{q}\ra X \piz$, and these events
are separated from the signal using the kinematic differences
between the signal and background.  The $\mes$ measured for the
four modes is shown in figure~\ref{fig:kstgam}, and the branching
fractions are summarized in table~\ref{tab:kstgam}.
\begin{table}
\caption{$\B \to \Kstar \gamma$ branching fractions.   Also given are the efficiency and the
number of signal events.}
\label{tab:kstgam}
\resizebox{0.45\textwidth}{!}{
\begin{tabular}{lccc} \hline
Mode        & $\epsilon$ (\%) &  Signal         &$\BR(\bkg)\times 10^{-5}$ \\ \hline\hline
a) $\Kp\pim$   &  14.1      & 135.7 $\pm$ 13.3  &     $4.39 \pm 0.41 \pm 0.27$   \\ 
b) $\Kp\piz$   &   5.1      &  57.6 $\pm$ 10.4  &    $5.52 \pm 1.07 \pm 0.33$   \\ 
c) $\KS\piz$   &   1.4      &  14.8 $\pm$  5.6  &    $4.10 \pm 1.71 \pm 0.42$   \\
d) $\KS\pip$   &   2.9      &  28.4 $\pm$  6.4  &    $3.12 \pm 0.76 \pm 0.21$   \\ \hline
\end{tabular}
}
\end{table}

\subsection{$B\rightarrow\phi K$and $B\rightarrow\phi K^*$}
\label{subsec:rarephik}
\begin{figure}
\resizebox{0.45\textwidth}{!}{%
\includegraphics{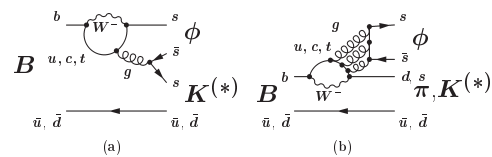}}
\caption{Penguin diagrams contributing to $B\rightarrow\phi K$ and $B\rightarrow\phi K^*$.}
\label{fig:phikfey}
\end{figure}
The decays $B\rightarrow\phi K$ and $B\rightarrow\phi K^*$ proceed primarily through
gluonic penguin diagrams in the Standard Model (figure~\ref{fig:phikfey}) and so
are expected to be sensitve to possible direct CP violating effects.  This mode also
provides the opportunity for a measurement of $\stwob$ which is complementary to that from 
the charmonium modes.

\begin{figure}
\resizebox{0.45\textwidth}{!}{%
\includegraphics{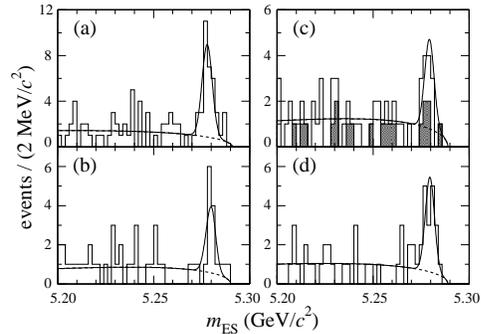}}
\caption{$\mes$ projections for $B\rightarrow\phi K$and $B\rightarrow\phi K^*$ decay modes.}
\label{fig:phikmes}
\end{figure}

The analysis of this channel\cite{phik} takes advantage of the excellent kaon ID in
\babar\ for high momentum kaons in order to reduce backgrounds.  A maximum likelihood
fit is performed using $\mes$, $\Delta_E$, and $M_{KK}$ as well as measurements of the particle ID
in the DIRC, and kinematic discriminants.   
Results of the fits are given in table~\ref{tab:phik}.  Signals have been seen in four modes,
including the first observations of $\phi K^{*+}$ and $\phi K^0$.
The projection of $\mes$ is shown in figure~\ref{fig:phikmes} for these modes.

\begin{table}
\caption{Measured branching fractions for $B\rightarrow\phi K$and $B\rightarrow\phi K^*$ decay modes.  Also
shown are the efficiency, the number of signal events, and the significance of the result.}
\label{tab:phik}
\resizebox{0.5\textwidth}{!}{
\begin{tabular}{lccccc}
\hline
\hline
Mode & $\varepsilon$  
& $n_{\mathrm{sig}}$ & $S$ & ${\cal B}(10^{-6})$ \cr
\hline
(a) $\phi K^+$   & 17.9 & $31.4^{+6.7}_{-5.9}$ & 10.5 &   $7.7^{+1.6}_{-1.4}\pm 0.8$  \cr
(b) $\phi K^0$    & 6.1  & $10.8^{+4.1}_{-3.3}$ &  6.4 &   $8.1^{+3.1}_{-2.5}\pm 0.8$  \cr
(c) $\phi K^{*+}$  & 4.9   & --                   &  4.5 &   $9.7^{+4.2}_{-3.4}\pm 1.7$  \cr
(d) $\phi K^{*0}$        & 8.6& $16.9^{+5.5}_{-4.7}$ & 6.6 & $8.6^{+2.8}_{-2.4}\pm 1.1$ \\\hline
$\phi \pi^+$          & 19.1 & $0.9^{+2.1}_{-0.9}$  & 0.6 & $<1.4$ (90\% CL)      \cr
\hline
\hline
\end{tabular}
}
\end{table}
\section{Semileptonic B Decays}
The large samples of fully reconstructed B decays at the B factories
allow us to explore new methods of reducing systematic uncertainties in
measuring semileptonic decay rates.  In the \babar\ Run~1 data sample there
are approximately 14,000 fully reconstructed B decays in about equal
amounts of charged and neutral B states.  For the charged B the largest
branching fraction modes are $\Bz \ra$
$\DorDstarm \pip$, $\DorDstarm \rho^+$,
 $\DorDstarm a_1^+$, 
$\jpsi  \Kstarz$ while for the neutral B 
$\Bu \ra$
$\DorDstarzb \pip$, $\jpsi K^+$,
 $\psitwos K^+ $ provide the largest number of events.  Once a B is fully
reconstructed, it is possible to determine whether the lepton found in
the other B in the event was a prompt lepton or a cascade lepton.  This
gives an independent measure of the prompt and cascase momentum
spectra. The overall number of prompt and cascade events for charged
B decays is given by $N^+_{right-sign}=N^+_p$, $N^+_{wrong-sign}=N^+_c$.
For neutral B decays, there is a dilution due to the mixing, $\chi_d$, which
is accomodated as
$N^0_{right-sign}=N^0_p(1-\chi_d)+N_c^0\chi_d$,
$N^0_{wrong-sign}=N^0_p\chi_d+N_c^0(1-\chi_d)$.

\begin{figure}
\resizebox{0.45\textwidth}{!}{%
\includegraphics{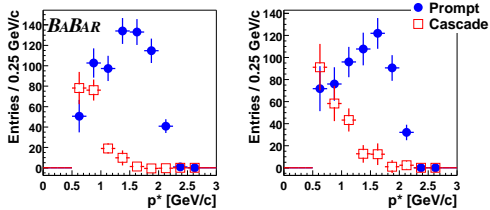}}
\caption{Momentum spectra for prompt and cascade leptons for $\Bz$ (left) 
and $\Bp$ (right) decays.}
\label{fig:semil}
\end{figure}

Figure~\ref{fig:semil} shows the momentum spectra extracted from the Run~1
data.  The measured branching fractions are
${\cal B}(B^+ \ra e^- X) = (10.3 \pm 0.6 \pm 0.5) \% $ and
${\cal B}(B^0 \ra e^- X) = (10.4 \pm 0.8 \pm 0.5) \% $
which give
$$ {\cal B}(B \ra e^- X) = (10.4 \pm 0.5 \pm 0.4) \% $$
$$ \frac{{\cal B}(B^+ \ra e^- X)}{{\cal B}(B^0 \ra e^- X)} = (0.99 \pm 0.10 \pm 0.03) \% $$.
\section{B Lifetimes, Mixing, and Searches for Direct CP Violation}
\subsection{B Lifetimes}\label{subsec:lifetime}
\begin{figure}
\parbox{0.45\textwidth}{
\begin{center}
\resizebox{0.4\textwidth}{!}{
      \includegraphics{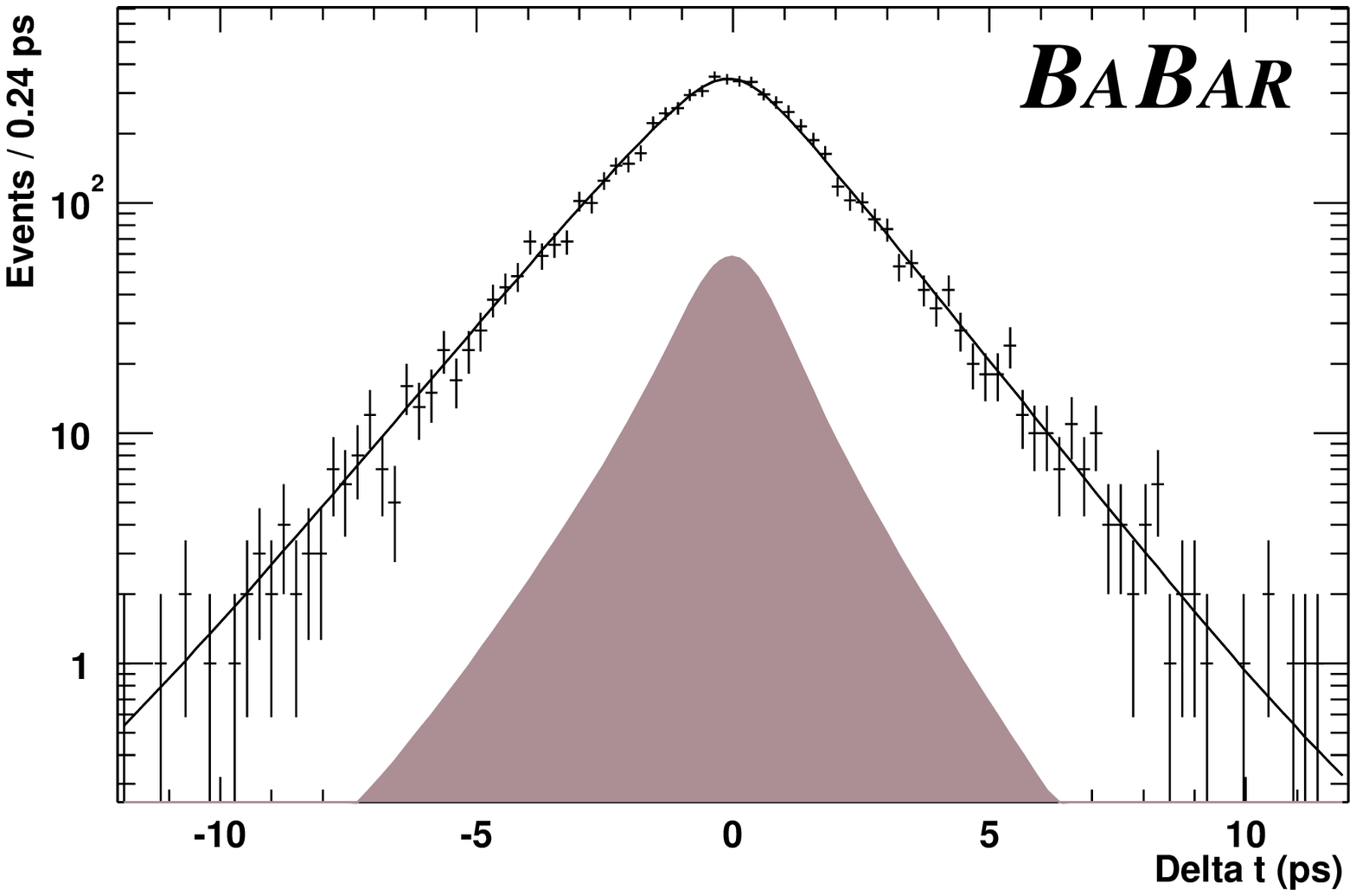}}
\resizebox{0.4\textwidth}{!}{
      \includegraphics{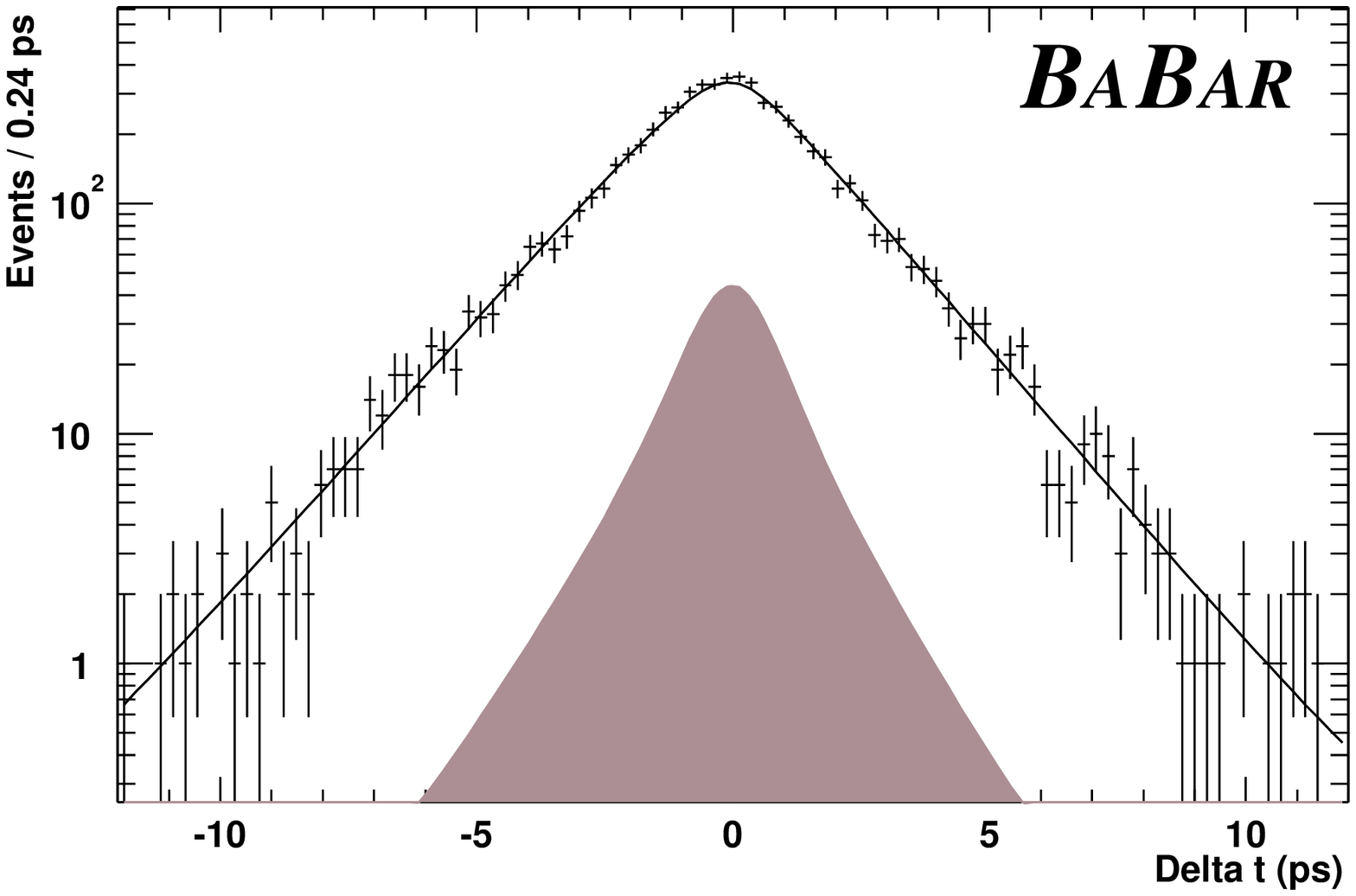}}
\end{center}
}
\caption{$\Delta t$ for $B^0$ (top) and $B^+$ (bottom).  The shaded area indicates the background.}
\label{fig:blife}
\end{figure}

Using fully reconstructed B decays also gives a substantial reduction
on the error in determining the $\Bz$ and $\Bp$ lifetimes.  The B factories
require new techniques in order to extract the B lifetimes
as the centre-of-mass
is boosted in the lab frame and there is no knowledge of the production
point of the Bs.  Instead one needs to measure the difference in flight
length which is directly sensitive to the lifetime.  \babar\ uses
the sample of fully reconstructed B decays with which one can vertex
and tag one B as either $\Bz$ or $\Bp$.  The tracks in the event not
associated with the fully reconstructed B are inclusively vertexed
to form the estimated decay point of the other B.  The knowledge of
the beamspot position is used to improve this vertex.  The width of the
distribution of the decay times differences is a combination of the
detector resolution and the B lifetimes.  These distributions are shown
in figure~\ref{fig:blife}, and are fit simultaneously in order to extract
the $\Bz$ and $\Bp$ lifetimes.\cite{blife}  The results of the fit are
$$  \tau_{\Bz} = 1.546 \pm 0.032\pm 0.022 \mbox{ ps,}$$
$$  \tau_{\Bu} = 1.673 \pm 0.032\pm 0.023 \mbox{ ps,}$$
$$   \tau_{\Bu}/\tau_{\Bz} = 1.082 \pm 0.026\pm 0.012$$
which are the best single measurements of these lifetimes.
\subsection{B Mixing with Leptons}\label{subsec:mixing}
\begin{figure}
\resizebox{0.45\textwidth}{!}{%
\includegraphics{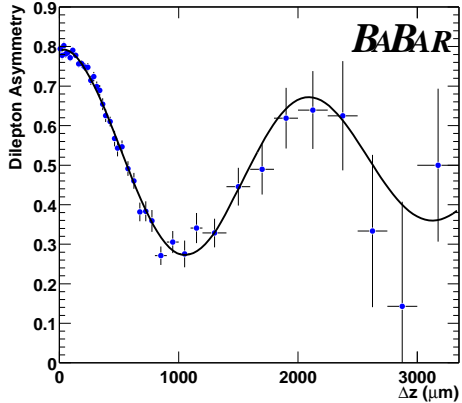}}
\caption{Time-dependent asymmetry for dilepton events $A(\Delta t)$.}
\label{fig:dilepmix}
\end{figure}

Events in which both Bs decay semi-leptonically allow one to measure the
B mixing parameter $\Delta m_d$ by measuring the time dependent
difference in the like sign vs unlike sign lepton events.  Measuring this
asymmetry
$$ A(\Delta t)   = \frac{N(\ell^+ \ell^-)(\Delta t) - N(\ell^\pm \ell^\pm)(\Delta t)}
                      {N(\ell^+ \ell^-)(\Delta t) + N(\ell^\pm \ell^\pm)(\Delta t)}$$
we extract\cite{slmix}
$\Delta m_d = 0.499\pm 0.010\pm 0.012 \,\hbar\ps^{-1}$ from the data shown
in figure~\ref{fig:dilepmix}.

\begin{figure}
\resizebox{0.45\textwidth}{!}{%
\includegraphics{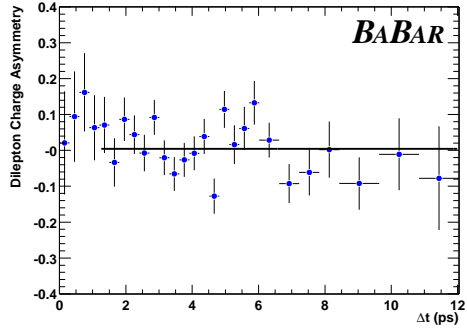}}
\caption{Time-dependent dilepton charge asymmetry $A_t(\Delta t)$.}
\label{fig:dilepcp}
\end{figure}

It is also possible to search for CP/T violation in mixing, measuring
$\epsilon_B$ from the dilepton charge asymmetry
$$ A_t(\Delta t) = \frac{N(\ell^+ \ell^+)(\Delta t) -N(\ell^-\ell^-)(\Delta t)}
                     {N(\ell^+\ell^+)(\Delta t)+N(\ell^-\ell^-)(\Delta t)} $$
$$\approx \frac{4{\mathrm Re}(\epsilon_B)}{1+|\epsilon_B|^2}$$
$\epsilon_B$ is the equivalent in the B system to the parameter $\epsilon$ in
the K system.  The data for $A_t(\Delta t)$ are
shown in figure~\ref{fig:dilepcp}. Measuring $A_t(\Delta t)$ we find \cite{cpt}
$$\frac{{\mathrm Re}(\epsilon_B)}{1+|\epsilon_B|^2}=(0.12\pm0.29\pm0.36)\%$$
which is the most stringent test of CP violation in B mixing.
\subsection{Searches for Direct CP Violation}\label{subsec:directCP}
Direct CP violation can be observed if there is a difference in both the
weak and strong phases between two different diagrams to the same
final state.   This effect can be searched for by looking for
a charge asymmetry in the observed final states.  One forms the
asymmetry
$${\cal A_{CP}} 
= \frac{ {\cal B}(\overline{B}\ra\overline{f}) - {\cal B}(B\ra f) } 
{ {\cal B}(\overline{B}\ra\overline{f}) + {\cal B}(B\ra f) }$$
$\approx |A_1||A_2|\sin\Delta\phi_W\sin\Delta\phi_S$
which is sensitive to any direct CP violating effects.  Currently
we see no evidence for direct CP violation\cite{directCP}, and a summary of the
limits we set are shown in table~\ref{tab:directCP}.
\begin{table}
\caption{summary of limits on direct CP violation in modes studied by \babar.}
\label{tab:directCP}
\begin{eqnarray*}
{\cal A_{CP}} (\eta' K^\pm) & = & -0.11\pm 0.11\pm 0.02\\
{\cal A_{CP}} (\omega\pi^\pm) & = & -0.01^{+0.29}_{-0.31}\pm 0.03\\
{\cal A_{CP}} (\phi K^\pm) & = & -0.05\pm 0.20\pm 0.03\\
{\cal A_{CP}} (\phi K^{*\pm}) & = & -0.43^{+0.36}_{-0.30}\pm 0.06\\
{\cal A_{CP}} (\phi K^{*0}) & = & 0.00\pm 0.27\pm 0.03\\
\Kstar\gamma \\
{\cal A_{CP}} (\Kpm\pi^\mp\gamma) & = & -0.035\pm 0.094\pm 0.022\\
{\cal A_{CP}} (\KS\pi^\pm\gamma) & = & -0.19\pm 0.21\pm 0.012\\
{\cal A_{CP}} (\Kpm\pi^0\gamma) & = & 0.044\pm 0.155\pm 0.021\\
{\cal A_{CP}} (\Kstar\gamma) & = & -0.035\pm 0.076\pm 0.012\\
\jpsi\Kpm \\
{\cal A_{CP}} (\jpsi \Kpm)  &=&  0.004\pm 0.029\pm 0.004\\
\end{eqnarray*}
\end{table}

\section{Conclusions}
The first Run of \babar\ has provided a glimpse at the potential
of the B factories for providing detailed tests of the CKM sector
of the Standard Model as well as probing for possible effects beyond
those predicted by the Standard Model.  The second Run of BaBar is
now undereway and will continue until July 2002.  By the end of this
Run it is anticipated that \babar\ will have recorded 100/fb of data
providing a rich sample to continue the search for rare B decays, and
measure CP violating effects in a variety of B decay modes.  As 
can be seen by the impressive results shown by our
colleagues at KEK today\cite{belle}, we expect a
healthy competition in compiling these results.
\section*{Acknowledgments}
It is a pleasure to acknowledge the conference organizers for the
delightful meeting in Rome.  I would also like to acknowledge our
colleagues at PEP II for providing the outstanding machine performance
which enabled us to produce these results.

\end{document}

%% file: symbolspsikst.tex
\def\az    {\ensuremath{A_{0}}}
\def\ap    {\ensuremath{A_{\parallel}}}
\def\at    {\ensuremath{A_{\perp}}}

\def\azd   {\ensuremath{|\az|^{2}}}
\def\apd   {\ensuremath{|\ap|^{2}}}
\def\atd   {\ensuremath{|\at|^{2}}}

\def\phip  {\ensuremath{\phi_{\parallel}}}
\def\phit  {\ensuremath{\phi_{\perp}}}



%% file: DDsymbols.tex
\def\BRbztodstdstNum {\ensuremath{(8.0 \pm 1.6 \pm
1.2)\times 10^{-4}} \xspace}
\def\BRbztodstdst {\ensuremath{{\BR}(\Bztodstdst) = \BRbztodstdstNum}\xspace}

\def\Dstarp     {\ensuremath{D^{*+}}\xspace}
\def\Dstarm     {\ensuremath{D^{*-}}\xspace}
\def\Dp         {\ensuremath{D^+}\xspace}
\def\Dm         {\ensuremath{D^-}\xspace}

\def\BtoDsDbar	{\ensuremath{B \to D^{(*)}_S \Dbar^{(*)}}\xspace}

\def\Dstptopip  {\ensuremath{\Dstarp \to \Dz\pip}\xspace}
\def\Dstptopiz  {\ensuremath{\Dstarp \to \Dp\piz}\xspace}

\def\DeltaE     {\ensuremath{\Delta E} \xspace}

%% file: Definitions.tex
\def\dbline{\noalign{\vskip 0.15truecm\hrule}\noalign{\vskip 2pt}\noalign{\hrule\vskip 0.15truecm}}
\def\sgline{\noalign{\vskip 0.20truecm\hrule\vskip 0.20truecm}}

\newcommand{\calB}{\mbox{${\cal B}$}}
\newcommand{\DE}{\ensuremath{\Delta E}}

\newcommand{\fetaKstp}{\mbox{$\eta K^{*+}$}}

\newcommand{\fetaKstz}{\mbox{$\eta K^{*0}$}}
\newcommand{\BetaKstp}{\mbox{$\calB(B^+\ra\eta K^{*+})$}}
\newcommand{\BetaKstz}{\mbox{$\calB(B^0\ra\eta K^{*0})$}}


%% file: nash_babar.bbl
\begin{thebibliography}{99}

\bibitem{bbrnim}
B.~Aubert {\it et al.}  [BABAR Collaboration],
arXiv:hep-ex/0105044.  To be published in NIM.

\bibitem{dorfan}
J. Dorfan, these proceedings.

\bibitem{sin2bprl}
B.~Aubert {\it et al.}  [BABAR Collaboration],
Phys.\ Rev.\ Lett.\  {\bf 87} (2001) 091801
[arXiv:hep-ex/0107013].

\bibitem{olsen}
S. Olsen, these proceedings.

\bibitem{psicont}
B.~Aubert {\it et al.}  [BABAR Collaboration],
Phys.\ Rev.\ Lett.\  {\bf 87} (2001) 162002
[arXiv:hep-ex/0106044].

\bibitem{exclcharm}
B.~Aubert {\it et al.}  [BABAR Collaboration],
arXiv:hep-ex/0107025.

\bibitem{psikpsipi}
B.~Aubert {\it et al.}  [BABAR Collaboration],
arXiv:hep-ex/0108009.

\bibitem{psikstar}
B.~Aubert {\it et al.}  [BABAR Collaboration],
Phys.\ Rev.\ Lett.\  {\bf 87} (2001) 241801
[arXiv:hep-ex/0107049].

\bibitem{cleoddk} CLEO Collaboration, CLEO~CONF~97-26, EPS97~337.

\bibitem{alephddk} ALEPH Collaboration, R.~Barate \emph{et al.}, 
Eur.~Phys.~J.{\bf C4}, 387-407 (1998).

\bibitem{buchalla} G.~Buchalla, I.~Dunietz and H.~Yamamoto, 
Phys.~Lett. {\bf B364}, 188 (1995). 

\bibitem{bddk}
B.~Aubert {\it et al.}  [BABAR Collaboration],
arXiv:hep-ex/0107056.

\bibitem{bdstdst}
B.~Aubert {\it et al.}  [BABAR Collaboration],
arXiv:hep-ex/0107057.

\bibitem{belle}
H. Tajima, these proceedings.

\bibitem{cl2b}
B.~Aubert {\it et al.}  [BABAR Collaboration],
Phys.\ Rev.\ Lett.\  {\bf 87} (2001) 151802
[arXiv:hep-ex/0105061],
B.~Aubert {\it et al.}  [BABAR Collaboration],
arXiv:hep-ex/0109005.

\bibitem{cleoetapk}
CLEO Collaboration, S.\ J.\ Richichi {\it et al.} \jprl{85}, 520 (2000);
CLEO CONF 99-12 (1999).

\bibitem{etapk}
B.~Aubert {\it et al.}  [BABAR Collaboration],
Phys.\ Rev.\ Lett.\  {\bf 87} (2001) 221802
[arXiv:hep-ex/0108017].

\bibitem{cleoetakst}
CLEO Collaboration S.J.\ Richichi {\it et al.}, \jprl{85}, 520 (2000); \\
CLEO CONF 99-12 (1999).
%
\bibitem{thy}
A. Ali, G. Kramer, and C.D. L\"{u}, \jprd {\textbf 58}, 094009 (1998); \\
Y. H. Chen {\it et al.}, \jprd {\textbf 60}, 094014 (1999).

\bibitem{etakst}
B.~Aubert {\it et al.}  [BABAR Collaboration],
arXiv:hep-ex/0107037.

\bibitem{bbrphys}
P.~F.~Harrison and H.~R.~Quinn  [BABAR Collaboration],
SLAC-R-0504.

\bibitem{a0pi}
B.~Aubert {\it et al.}  [BABAR Collaboration],
arXiv:hep-ex/0107075.

\bibitem{rhopi}
B.~Aubert {\it et al.}  [BABAR Collaboration],
arXiv:hep-ex/0107058.

\bibitem{bgamgam}
B.~Aubert {\it et al.}  [BABAR Collaboration],
Phys.\ Rev.\ Lett.\  {\bf 87} (2001) 241803
[arXiv:hep-ex/0107068].

\bibitem{kllth}
{A.~Ali, P.~Ball, L.T.~Handoko, and G.~Hiller, Phys.~Rev.~D
{\bf 61}, 074024 (2000); hep-ph/9910221.}\\
{T.M.~Aliev, A.~Ozpineci and M.~Savci, Phys.~Rev.~D {\bf 56}, 4260 (1997);
hep-ph/9612480.}\\
{T.M.~Aliev, C.S.~Kim, and Y.G.~Kim, Phys.~Rev. ~D {\bf 62}, 014026
(2000); hep-ph/9910501.}\\
{D.~Melikhov, N.~Nikitin, and S.~Simula, Phys.~Lett. B{\bf 410}, 290 (1997);
hep-ph/9704628.}\\
{D.~Melikhov and B.~Stech, Phys.~Rev.~D {\bf 62}, 014006 (2000); 
hep-ph/0001113.}\\
{P.~Colangelo, F.~De Fazio, P.~Santorelli, and E.~Scrimieri,
Phys.~Rev.~D {\bf 53}, 3672 (1996); Erratum--{\it ibid.} D{\bf 57}, 3186 (1998); hep-ph/951043.}\\
{P.~Colangelo {\it et al.}, Eur.~Phys.~J. C{\bf 8}, 81 (1999);
hep-ph/9809372.}

\bibitem{kll}
B.~Aubert {\it et al.}  [BABAR Collaboration],
arXiv:hep-ex/0107026.

\bibitem{kstargam}
B.~Aubert {\it et al.}  [BABAR Collaboration],
arXiv:hep-ex/0110065.

\bibitem{phik}
B.~Aubert {\it et al.}  [BABAR Collaboration],
Phys.\ Rev.\ Lett.\  {\bf 87} (2001) 151801
[arXiv:hep-ex/0105001].

\bibitem{blife}
B.~Aubert {\it et al.}  [BABAR Collaboration],
Phys.\ Rev.\ Lett.\  {\bf 87} (2001) 201803
[arXiv:hep-ex/0107019].

\bibitem{slmix}
B.~Aubert {\it et al.}  [BABAR Collaboration],
arXiv:hep-ex/0008054.

\bibitem{directCP}
B.~Aubert {\it et al.}  [BABAR Collaboration],
arXiv:hep-ex/0109006.

\bibitem{cpt}
B.~Aubert {\it et al.}  [BABAR Collaboration],
arXiv:hep-ex/0107059.

\end{thebibliography}
